\documentclass{aa}

\usepackage{epsfig,graphicx,color} \usepackage {txfonts } \usepackage
           {natbib } \usepackage {url }

\begin{document}

%
%

\title {Worldwide site comparison for submillimetre astronomy
       }


\author{ P. Tremblin \inst{1}\and N. Schneider \inst{1,2}\and V. Minier
  \inst{1}\and G. Al. Durand \inst{1}\and J. Urban \inst{3} }

\institute{Laboratoire AIM Paris-Saclay (CEA/Irfu - Uni. Paris Diderot
  - CNRS/INSU), Centre d'\'etudes de Saclay, 91191 Gif-Sur-Yvette,
  France  
\and Universit\'e de Bordeaux1, LAB, UMR 5804, CNRS, 33270 Floirac, France
\and Chalmers University of Technology, Department of Earth
  and Space Sciences, 41296 G\"oteborg, Sweden
}

\date{
} \offprints{P. Tremblin} \mail{pascal.tremblin@cea.fr}

\titlerunning{Worldwide site comparison} \authorrunning{Tremblin et al.}

\abstract{}
{The most important limitation for ground-based submillimetre (submm)
  astronomy is the broad-band absorption of the total water vapour in
  the atmosphere above an observation site, often
  expressed as the Precipitable Water Vapour (PWV). A long-term statistic
  on the PWV is thus mandatory to characterize the quality of an
  existing or potential site for observational submm-astronomy.  In
  this study we present a three-year statistic (2008--2010) of the
  PWV for ground-based telescope sites all around the world and for
  stratospheric altitudes relevant for SOFIA (Stratospheric
  Observatory for Far-infrared astronomy). The submm-transmission is
   calculated for typical PWVs using an atmospheric model.}
%
{We used data from IASI (Infrared Atmospheric Sounding Interferometer)
  on the Metop-A satellite to retrieve water vapour profiles for each site (11 in
  total, comprising Antarctica, Chile, Mauna Kea, Greenland, Tibet).
  The use of a single instrument to make the comparison provides
  unbiased data with a common calibration method. The profiles are
   integrated above the mountain/stratospheric altitude
  to get an estimation of the PWV. We then applied the atmospheric model
  MOLIERE (Microwave Observation and LIne Estimation and REtrieval) to
  compute the corresponding atmospheric absorption for wavelengths
  between 150 $\mu$m and 3 mm.}
%
{We present the absolute PWV values for each site sorted by year and
  time percentage. The PWV corresponding to the first decile (10\%)
  and the quartiles (25\%, 50\%, 75\%) are calculated and transmission
  curves between 150 $\mu$m and 3 mm for these values are
  shown. The Antarctic and South-American sites present
  very good conditions for submillimetre astronomy. The 350 $\mu$m and
  450 $\mu$m atmospheric windows are open all year long whereas the
  200 $\mu$m atmospheric window opens reasonably for 25 \% of the time
  in Antarctica and the extremely high-altitude sites in
  Chile. Potential interesting new facilities are Macon in Argentinia
  and Summit in Greenland that show similar conditions as for example
  Mauna Kea (Hawaii).  For SOFIA, we present in more detail
  transmission curves for different altitudes (11 to 14 km), PWV
  values, and higher frequencies (up to 5 THz).  Though the atmosphere
  at these altitude is generally very transparent, the absorption at very
  high frequencies becomes more important, partly caused by minor
  species. In conclusion, the method presented in this paper could identify sites on
    Earth with a great potential for submillimetre astronomy, and
    guide future site testing campaigns \it{in situ}. }
{}


\maketitle

%
%

\section{Introduction}


Despite its large interest for astronomy, the submillimetre (submm) to
far-infrared (FIR) wavelength range is mostly limited by the
pressure-broadened absorption of tropospheric water vapour present in
the Earth's atmosphere \citep[see, e.g.,][for a summary of astronomy
  projects]{Minier:2010dg}. The water vapour profile,
and thus 
also the integrated column of water expressed in Precipitable Water
Vapour (PWV), depends on the geographical location. The PWV generally
decreases with altitude, so that high-lying sites are best suited for
submm-astronomy. Airborne astronomy, now possible on a regular base
using SOFIA, is much less limited by water vapour and due to the lower
atmospheric pressure, spectral broadening is less important. However,
there is still absorption close to major and minor species (e.g.,
O$_3$, N$_2$O, CO) as well as collision induced non-resonant continuum
absorption. SOFIA can now access the very high frequency range ($>$2
THz), definitely closed for ground-based sites, but line intensities
and abundances of the major and minor species need to be known
precisely. First spectroscopic observations performed with SOFIA (see
A\&A special issue) showed that small-scale atmospheric variations in
the troposphere/stratosphere layer are very difficult to assess
\citep{Guan:2012ce}. 

For ground-based sites, previous studies
\citep[e.g.][]{Schneider:2009hz,Tremblin:2011fg,Matsushita:1999wh,Peterson:2003dq}
already showed that a few 
sites are well-suited for submm/mid-IR, and FIR-astronomy and their
transmission properties are rather well determined (for example by
Fourier Transform Spectrometer obervations in the 0.5--1.6 THz range
 at Mauna Kea/Hawaii \citep{Pardo:2001km}). The
high-altitude ($>$5000 m) Chilean sites are known for dry conditions
\citep[see][]{Matsushita:1999wh,Peterson:2003dq}, and site testing is
now carried out at the driest place on Earth, Antarctica
\citep[see][]{Chamberlin:1997jc,Yang:2010jg,Tremblin:2011fg}.
Comparisons between Antarctic and Chilean sites are difficult and
uncertain since they rely on ground-based instruments that use
different methods and calibration techniques \citep[see][for
  example]{Peterson:2003dq}. The working conditions are also an
important issue, a single instrument moved from Chile to Antarctica
will have a different behavior in the harsh polar environment (
-70$^\circ$C in winter at Dome C). A meaningful comparison is possible if several independent instruments
are used at each place. An example of such a study is the one of
\citet{Tremblin:2011fg} that obtained transmission data at Dome C
thanks to radio-soundings and the radiometers HAMSTRAD \citep{Ricaud:2010ik} and
SUMMIT08. However, it is rare to have many instruments at
one site. The best solution is to use satellite data, which also
enables to investigate any location on Earth. Thanks to the IASI
(Infrared Atmospheric Sounding Interferometer) on the Metop-A satellite, it is now
possible to conduct such a comparison over several years with no
instrument bias and with the same working conditions for the
detectors. We present here a 3-year study of the PWV of a selection of
existing and upcoming submm-sites in Antarctica, Chile, Tibet, and
Argentina, as well as for two SOFIA stations, Palmdale/California and
Christchurch/New Zealand. 
The individual and cumulated quartiles of PWV for each site allow a
direct comparison. The transmission corresponding to these PWV values
of the quartiles is then calculated using the atmospheric model
MOLIERE-5 \citep{Urban:2004cr}. 
The paper is organized as follows: After
introducing all the sites that are considered in the present study
(Sect.~2), we first show the PWV extraction from the satellite data and
the comparison of the PWV statistics between all the sites
(Sect.~3). Then we present the method to extract transmission curves
with the MOLIERE atmospheric model and perform a site comparison for
the 250 $\mu$m and 350 $\mu$m transmission (Sect.~4)  The ratio of the
monthly-averaged transmission 
to its fluctuations is then used to compare the stability of the
transmission at 200 $\mu$m between all the sites at the end of Sect.~4. Section 5
deals with the results for SOFIA and Sect.~6 discusses and summarizes
the results.


%
%

\section{Sites of interest}

We selected eleven representative sites around the world that can be
grouped as follows: \\

\noindent \underline{Antarctic sites}\\
\noindent {\sl Dome C:} Concordia station  with
several telescopes, e.g. IRAIT (International Robotic Antarctic
Infrared Telescope), http://arena.oca.eu \\
\noindent {\sl Dome A:} new site in exploration,
 http://www.chinare.gov.cn \\
\noindent {\sl South pole:} SPT (South Pole Telescope),
\\ http://astro.uchicago.edu/research/south-pole-telescope \\

\noindent \underline{South-American sites}\\
\noindent {\sl Chajnantor Plateau in Chile}: ALMA (Atacama Large
Millimetre Array) site, http://www.almaobservatory.org \\
\noindent {\sl Cerro Chajnantor in Chile:} CCAT (Cornell Caltech
Atacama Telescope), http://www.ccatobservatory.org \\
\noindent {\sl Cerro Macon in Argentinia} \\

\noindent \underline{Northern-hemisphere sites} \\
\noindent {\sl Mauna Kea in Hawaii:} JCMT (James Clerk Maxwell
Telescope), http://www.jach.hawaii.edu \\
\noindent {\sl Summit in Greenland:} http://www.geosummit.org \\
\noindent {\sl Yangbajing in Tibet:} KOSMA (Cologne Observatory for
submm-astronomy), http://www.astro.uni-koeln.de/kosma \\

\noindent \underline{``Stratospheric'' sites}\\
\noindent {\sl SOFIA} over Palmdale (USA) and Christchurch (New
Zealand), http://www.sofia.usra.edu \\

\begin{figure}[!ht]
\centering \includegraphics[width=\linewidth]{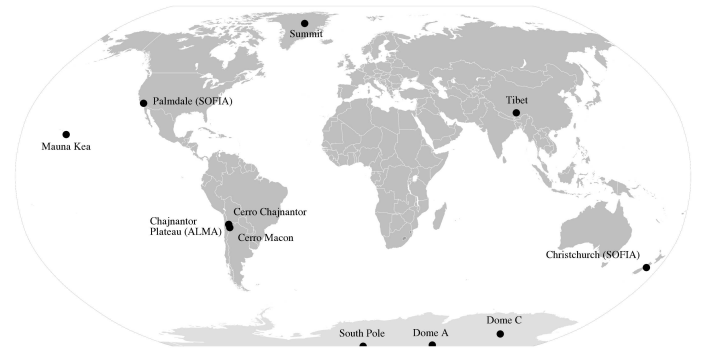}
\caption{Location of the different sites of interest.}
\label{world_map}
\end  {figure}

The location of the different sites are resumed in
Fig. \ref{world_map}. Our selection inclues known sites in Antarctica
and North/South America plus northern-hemispheric sites that could be
of interests (Summit in Greenland, Yangbajing in Tibet) to have a complete sky coverage. The
geographical location, altitude, and related existing/planned
telescopes are listed in Table \ref{sites}. \\



\begin{table}[!ht]
\caption{\label{sites} Location, altitude and telescopes for the
  different sites. Future telescopes are indicated in parenthesis.}
\centering
\begin{tabular}{lrrrr}
\hline \hline Site & Latitude & Longitude & Alt. [m] & Tel. \\ \hline
\hline Dome C & 75$^\circ$06$'$s & 123$^\circ$23$'$e & 3233 & IRAIT
\\ Dome A & 80$^\circ$22$'$s & 77$^\circ$21$'$e & 4083 & \\ South pole
& 90$^\circ$s & 0$^\circ$e & 2800 & SPT \\ \hline Cerro Chajnantor &
22$^\circ$59$'$s & 67$^\circ$45$'$w & 5612 & (CCAT) \\ Cerro Macon &
24$^\circ$31$'$s & 67$^\circ$21$'$w & 5032 & \\ Chaj. Plateau &
23$^\circ$00$'$s & 67$^\circ$45$'$w & 5100 & ALMA \\ \hline Mauna Kea
& 19$^\circ$45$'$n & 155$^\circ$27$'$w & 4207 & JCMT \\ Summit (Greenland)&
72$^\circ$35$'$n & 38$^\circ$25$'$w & 3210 & \\ Yangbajing (Tibet) &
30$^\circ$05$'$n & 90$^\circ$33$'$e & 4300 & KOSMA \\ \hline Palmdale
& 34$^\circ$37$'$n & 118$^\circ$05$'$w & 12000 & SOFIA \\ Christchurch
& 43$^\circ$31$'$s & 172$^\circ$38$'$e & 12000 & SOFIA \\
\end{tabular}
\end{table}

%
%

\section{PWV extraction from satellite data}

\subsection{Method and results} \label{method}

IASI (Infrared Atmospheric Sounding Interferometer) is an atmospheric
interferometer working in the infrared, launched in 2006 on the
METOP-A satellite \citep{Phulpin:2007dj,Pougatchev:2008jd,Herbin:2009vr,Clerbaux:2009ts}. The data
are available at the website of the Centre for Atmospheric Chemistry
Products and
Services\footnote{\url{http://ether.ipsl.jussieu.fr}}. IASI was
developed by the French Space Agency CNES\footnote{Centre National
  d'Etudes Spatiales} in collaboration with EUMETSAT\footnote{European
  organization for meteorological satellites
  \url{http://www.eumetsat.int}}. The satellite is on a polar orbit so
that each point on Earth is seen at least once per day by the
detectors. IASI is a Fourier Transform Spectrometer (FTS) working
between 3.7 and 15.5 $\mu$m. It is associated with an infrared imager,
operating between 10.3 and 12.5 $\mu$m. Each pixel of the instrument
corresponds to a spatial extent of 12 km at nadir, and vertical
profiles of tropospheric humidity at ninety altitude levels (resolution 1 km) are retrieved with a
typically 10\% accuracy \citep{Pougatchev:2008jd}. The amount of precipitable water vapour is given by the
integral of these vertical profiles. We averaged
on a daily base all measurements whose central position falls in a zone of 110 km$^2$ around each site to derive PWV
statistics between 2008 and 2010. This method was validated
in \citet{Tremblin:2011fg} in which we compared IASI satellite data with
ground-based instruments (HAMSTRAD and SUMIT08) and radio-sounding
data over the French-Italian base Concordia at Dome C in
Antarctica. All instruments have a good correlation at low PWV which
is precisely the range we are interested in for submillimetre
astronomy.

\begin{figure}[!ht]
\centering \includegraphics[width=\linewidth]{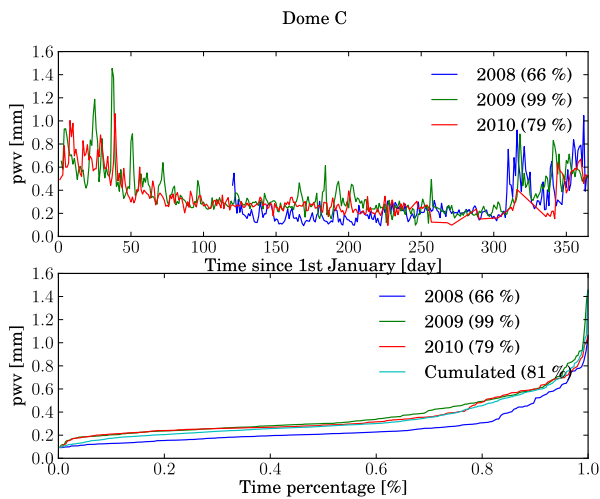}
\caption{Top: PWV content measured at Dome C, Antarctica over the
  French Italian base Concordia between 2008 and 2010. The percentage
  indicates the fraction of days of the year on which we were able to
  extract the data from the satellite measurements. Bottom: function
  of repartition of the PWV for each year and for the whole period.}
\label{domec_pwv}
\end  {figure}

\begin{figure}[!ht]
\centering \includegraphics[width=\linewidth]{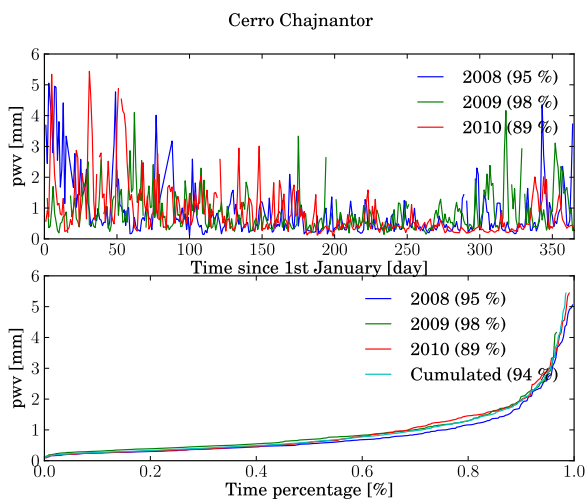}
\caption{Same as Fig. \ref{domec_pwv} for Cerro Chajnantor in
  Chile. The satellite profiles were
  truncated at the altitude of the mountain summit. Note the different
scale range of PWV compared to Fig. \ref{domec_pwv}.}
\label{ccat_pwv}
\end  {figure}

\begin{table*}[!ht]
\caption{\label{pwv_table} First decile and quartiles of the PWV for
  all the studied sites.}  \centering
\begin{tabular}{cccccc}
\hline Time fraction 2008-2010 & SOFIA (Palm./Christ.) & Dome A & Dome C & South Pole & Cerro Chajnantor \\ \hline 
0.10 & 0.006/0.004 & 0.11 & 0.17 & 0.15 & 0.27 \\ 
0.25 & 0.006/0.005 & 0.16 & 0.22 & 0.21 & 0.37 \\ 
0.50 & 0.007/0.006 & 0.21 & 0.28 & 0.30 & 0.61 \\ 
0.75 & 0.009/0.007 & 0.26 & 0.39 & 0.49 & 1.11 \\ 
\hline \hline Time fraction 
2008-2010 & Chajnantor Plateau & Summit & Cerro Macon & Mauna Kea & Yangbajing \\ \hline 
0.10 & 0.39 & 0.36 & 0.47 & 0.62 & 1.21 \\ 
0.25 & 0.53 & 0.51 & 0.66 & 0.91 & 2.47 \\ 
0.50 & 0.86 & 0.94 & 1.02 & 1.44 & inf \\ 
0.75 & 1.63 & 1.96 & 1.66 & 2.57 & inf \\
\end{tabular}
\end{table*}

The use of vertical satellite profiles is slightly trickier for a
mountain site. Since we take all measurements in a zone of 110 km$^2$,
we sometimes get profiles that do not contain mountain altitude but include
lower ones. This would bias the retrieved PWV to high values. To
overcome this difficulty, we generally truncated the profiles at the
pressure level of the site of interest. This method was already used by
\citet{Ricaud:2010ik} to compare IASI measurements with the HAMSTRAD
radiometer, over the Pyrenees mountains. They showed a very good
correlation for the integrated PWV. We also determined in this way the
PWV content at high altitudes ($>$11 km) over Palmdale, USA and
Christchurch, New Zealand for the on-going and future flights of
SOFIA. During a typical flight, SOFIA will range hundreds of
  kilometers from its base and the water vapour depends on local
  weather conditions at the position of the plane. Since the profiles
  are averaged on a 110-km$^2$ area and the time statistics is done for 
  three years, the results given here are representative 
  conditions under which SOFIA is operating. Precise in-situ 
  measurements of the PWV useable for calibration are much more
  difficult to obtain \citep[see][]{Guan:2012ce}.

The water vapour profiles are measured as a function of 
pressure by IASI. Therefore it is important to get the correct
pressure level at the altitude of the sites. In-situ measurements were
available for most of the sites to get the local averaged pressure
level. Only for Yangbajing, Cerro Macon, and SOFIA, it was neccessary
to use the pressure level of the US standard atmosphere (1976),
computed at the altitude of the site. However, this method does not
provide a precise pressure level just as a measured one so that the
retrieved PWVs are less reliable. By comparing both methods for sites
were measurements were available, we estimate an error of around 25~\%
using the standard atmosphere.


Figure~\ref{domec_pwv} presents the PWV measurements for the Antarctic
site of Dome C and Fig. \ref{ccat_pwv} the measurements for the
Chilean site of Cerro Chajnantor. The corresponding figures for all
other sites are found in the Appendix and on a dedicated
website\footnote{\url{http://submm.eu}}.
Since on Antarctic domes (expression for the mountains in Antarctica),
the difference in height is small over 100 km, we do not need to
truncate the profiles, whereas it has to be done for a mountain site
like Cerro Chajnantor. Some ground measurements are available for
different sites and can be compared to our results. Cerro Chajnantor
and Chajnantor Plateau were extensively studied
\citep[see][]{Giovanelli:2001gi} and the PWV quartiles
between 2006 and 2010 can be found in \citet{Radford:2011wq}. The
values deduced from opacity measurements at 225 GHz are 0.61 mm
(25 \%), 1.08 mm (50 \%), and 2.01 mm (75 \%) for the plateau; and
0.33 mm (25 \%), 0.61 mm (50 \%), and 1.36 mm (75 \%) for Cerro
Chajnantor. The quartiles for Cerro Chajnantor are in a good
agreement with our satellite measurements (between 10 and 20 \%
difference). Note that the studied periods are not the same and can
explain partly the difference. Furthermore, a scale height of the
water vapour profiles on the Chajnantor area can be
deduced from the quartiles of the plateau and the ``Cerro''. With the
satellite measurements, we computed a 1/e scale height $h_{e}$ of the
order of 1.1 km that is in a good agreement with measurements from
\citet{Giovanelli:2001gi}. The PWV quartiles at the South Pole were
computed at saturation from radio-sounding balloons over a period of 40 years
\citep[see][]{Chamberlin:2002tl}. These authors  estimated that the PWV should
be at 90 \% saturated, thus the saturation level can be
used as a proxy for the real PWV. They found averaged quartiles of
0.25 mm (25 \%), 0.33 mm (50 \%), and 0.44 mm (75 \%); the difference
with our satellite measurements is around 10 \%. A radiometer at 225
GHz  is operating at the CSO\footnote{Caltech Submillimeter
  Observatory} on Mauna Kea. PWV quartiles between 1997 and 2001 can
be found in \citet{Otarola:2010ht,Radford:2011wq}, 1.08 mm (25 \%),
1.82 mm (50 \%), and 3.32 mm (75 \%). The difference with the
satellite measurements is around 15-20 \% and can again be caused
by variations between the period 1997-2001 and 2008-2010.\\

The effect of seasons are clearly visible in 
Figs. \ref{domec_pwv} and \ref{ccat_pwv}. Obviously, the PWV during winter is lower than in
summer, i.e., below 0.3 mm at Dome C and below 0.6 mm at Cerro
Chajnantor. The Chilean site presents a strong day-to-day variation
compared to the Antarctic site. This is caused by the daily variation
of the PWV that is not present in Antarctica with the long polar
night/winter. This may cause a slight bias in the non-polar site data
because only one or two measurements per day are available, thus we do
not sample very well these day-night variations. Nevertheless, we
expect these statistics to represent well the trends over years. The
functions of repartition in Fig. \ref{ccat_pwv} are very similar from
2008 and 2010, therefore the year-to-year statistics are not too much affected
by the lack of the day-night variation sampling. In 2008 the PWV was
slightly lower than in 2009-2010. This
effect was seen in Antarctica (see Fig. \ref{domec_pwv}) also with the
in-situ measurements (radio-soundings, HAMSTRAD, SUMMIT08). Possible
explanations were discussed in \citet{Tremblin:2011fg} and should not
be interpreted as a possible effect of the day-night variations.

\subsection{Site comparison}

\begin{figure}[!ht]
\centering \includegraphics[width=\linewidth]{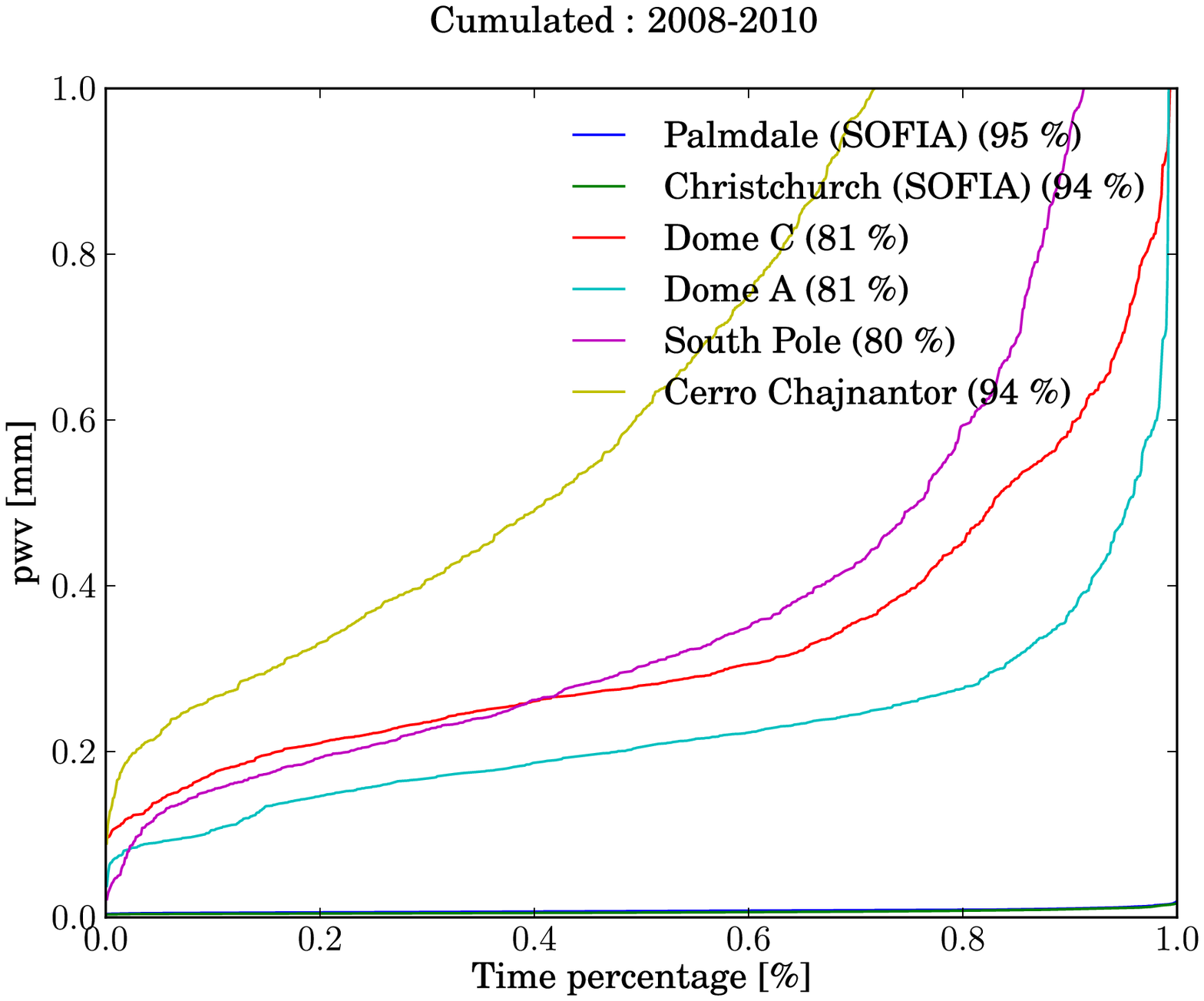}
\includegraphics[width=\linewidth]{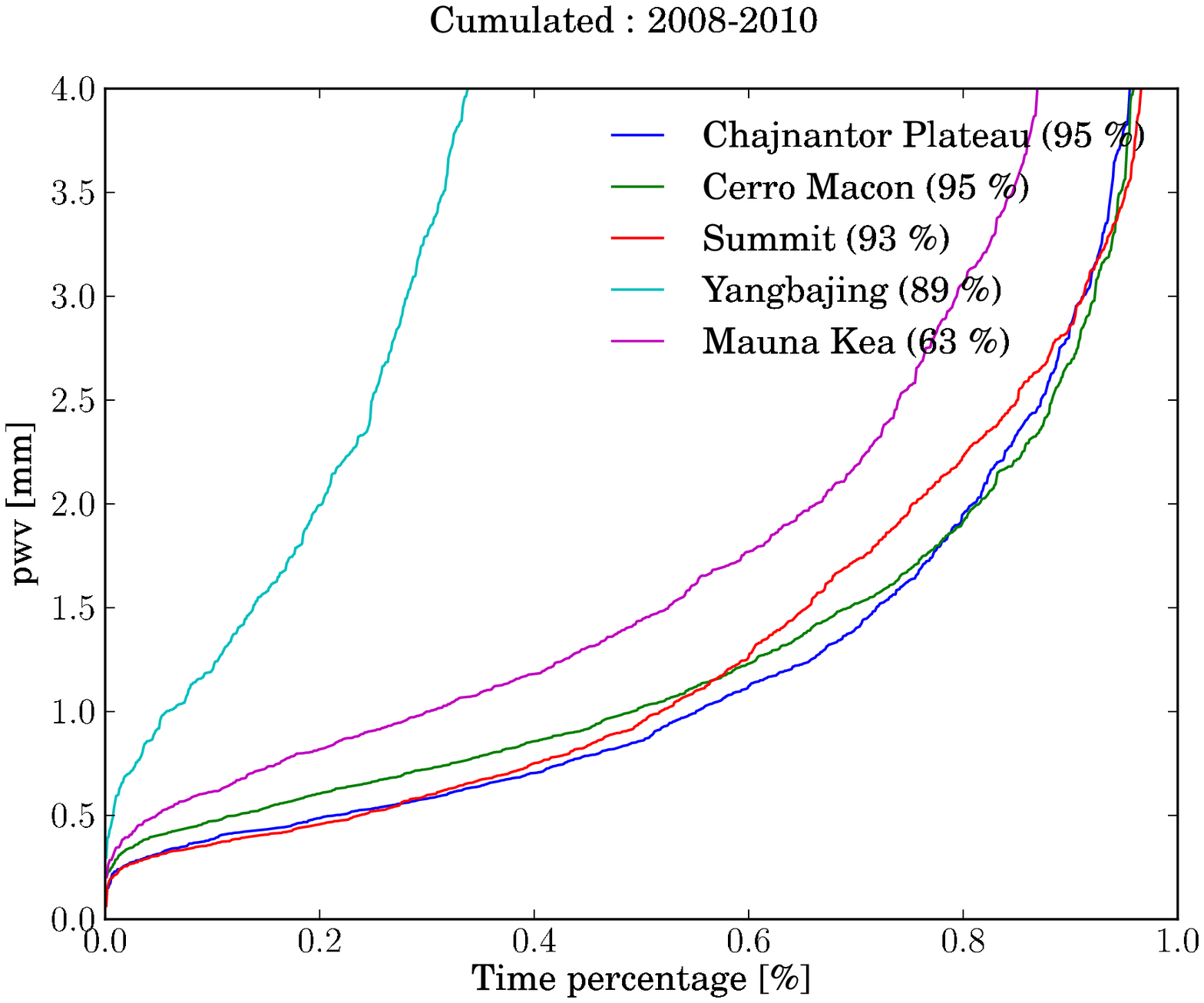}
\caption{Comparison of the functions of repartition of the PWV for all
  sites between 2008 and 2010. The percentage of days over this period
  on which we achieved the data extraction is indicated in parenthesis.
  The time percentage (x-axis) is made over the total number of days on
  which we achieved the data extraction. Note
  that in the upper panel, the 
  curves for Palmdale 
  (dark blue) and Christchurch (green) are blended and very close to
  the 0.0 mm level.}
\label{pwv_comp}
\end  {figure}

Figure~\ref{pwv_comp} shows the comparison between all sites as a
function of repartition of the PWV over the period
2008-2010. Obviously, the air-borne SOFIA stratospheric instrument encounters
the lowest PWV, lower than 0.01 mm most of the time and much lower
than any ground-based sites. The Antarctic sites follow then as the
driest places on Earth, with Dome A, and then Dome C and the South
Pole. These sites have PWVs lower than 0.5 mm most of the time (75 \% of the
time between 2008 and 2010). Chilean and Argentinian sites are next
with a PWV lower than 1.5 mm most of the time, followed by Hawaii and
Summit in Greenland with a PWV lower than 2 mm, and finally Yangbajing
in Tibet which presents 
a high level of water vapour. The satellite measurements
saturate over the Tibet site with high values of the PWV (around 10
mm), this is clearly visible in the figure in the
appendix.\\ These results are presented quantitatively in Table
\ref{pwv_table} with the first decile and the quartiles for each
site. The difference caused by the truncation for mountain sites can
be seen by comparing Cerro Chajnantor to Chajnantor Plateau. The
mountain (Cerro) is close to the ALMA site, therefore the profiles
that are extracted from the satellite are the same, the difference in
the PWV retrieved is mostly caused by the altitude truncation of the
profiles. The first decile and the quartiles again clearly show that
Antarctic sites are the driest sites followed by South-American sites
and then northern-hemisphere sites. It is remarkable that our long-term
satellite statistic of PWV shows that the site of Summit in Greenland
offers comparable observing conditions (PWV and altitude) like the ones 
on Mauna Kea which opens a new perspective for submm astronomy in the
northern hemisphere.

%
%

\section{Atmospheric transmission from PWV and MOLIERE}\label{transmission_sect}

\subsection{Method}

\begin{figure}[!ht]
\centering \includegraphics[width=0.6\linewidth,angle=-90]{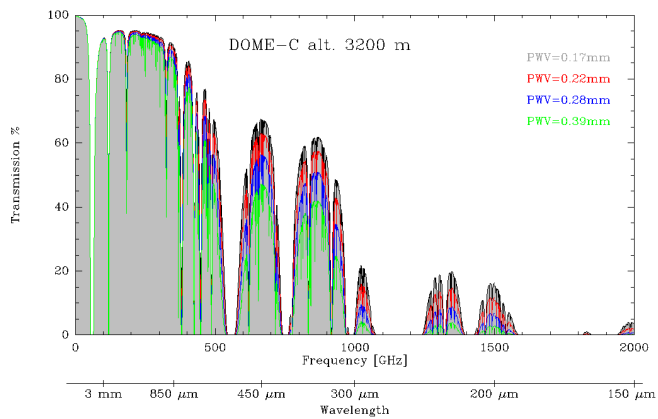}
\caption{Transmission curves between 150 $\mu$m and 3 mm corresponding
  to the first decile (grey) and the quartiles (red: 25 \%, blue: 50
  \%, green: 75 \%) of PWV between 2008 and 2010 at Dome C.}
\label{domec_moliere}
\end  {figure}

\begin{figure}[!ht]
\centering \includegraphics[width=0.6\linewidth,angle=-90]{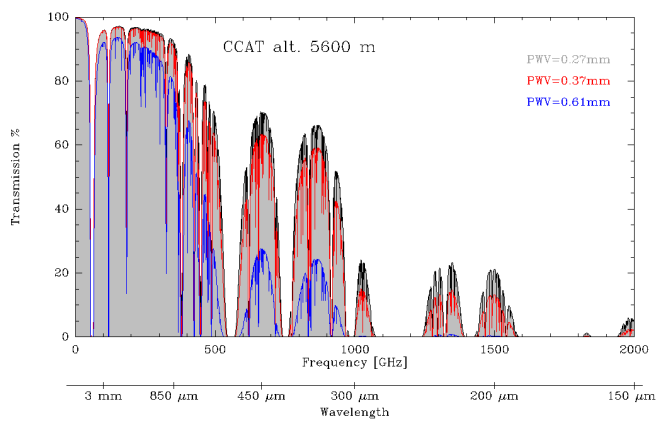}
\caption{Transmission curves between 150 $\mu$m and 3 mm corresponding
  to the first decile (grey) and the quartiles (red: 25 \%, blue: 50
  \%) of PWV between 2008 and 2010 at Cerro Chajnantor. }
\label{ccat_moliere}
\end  {figure}

For the determination of the tropospheric transmission corresponding
to the PWVs of the various deciles and quartiles for each site, we use
{\it MOLIERE-5.7} (Microwave Observation and LIne Estimation and
REtrieval), a forward and inversion atmospheric model
\citep{Urban:2004cr}, developed for atmospheric science
applications. It has previously been used to calculate the atmospheric
transmission up to 2000 GHz ($\sim$150 $\mu$m) for a large number of
astronomical sites. The results are published in
\citet{Schneider:2009hz} and on a
website\footnote{\url{http://submm.eu}}.

The code calculates the absorption of radiation in the mm- to far-IR
wavelength range (equivalent from 0 to 10 THz in frequency)
considering the wet- and dry-air components of atmospheric
absorption. Spectroscopic lines such as those of atmospheric water
(H$_2$O), oxygen (O$_2$), and ozone (O$_3$) absorb strongly at short
wavelengths, while collisions of H$_2$O with O$_2$ and nitrogen
(N$_2$) result in continuous absorption across all wavelengths. For
the line absorption, a radiative transfer model including refraction
and absorption by major and minor atmospheric species is included
(H$_2$O, O$_2$, O$_3$, NO$_2$, HNO$_3$, CO and other lines up to 10
THz). Spectroscopic parameters are taken from HITRAN\footnote{High
  Resolution Transmission molecular absorption database} and the
JPL\footnote{Jet Propulsion Laboratory} database.  The wet and dry air
continuum absorption is calculated as described in
\citet{Schneider:2009hz}. Temperature and pressure profiles of the
sites were 
determined using the compilation described in
\citet{Schneider:2009hz} and given on the website. The colder Antarctic atmosphere
preferentially populates the low-lying energy state of water, leading
to a greater absorption for a given PWV than that of a non-polar
site. This effect is included in the model.

We present here the transmission curves corresponding to the first
decile and the quartiles of PWV for two sites of currently large
interest, Dome C (Fig. \ref{domec_moliere}) and Cerro Chajnantor
(Fig. \ref{ccat_moliere}) while transmission curves for the other
sites are found in the appendix. From Table~\ref{pwv_table} we extract
that Dome C shows generally lower PWV values than Cerro Chajnator for
longer time periods. Comparing the transmission curves at the same PWV
(for example 0.28/0.27 mm which corresponds to 50 \% of the time at
Dome C (blue curve) and 10 \% of the time at Cerro Chajnator (grey
line)) indicates, however, that the transmission is slighhtly higher
at Cerro Chajnator. It corresponds to 20 (65) \% at 200 $\mu$m (350
$\mu$m) for Cerro Chajnator with regard to $\sim$10 (60) \% at 200
$\mu$m (350 $\mu$m) for Dome C. The difference becomes more important for high
frequencies (shorter wavelengths) because the absorption by water
vapour and the dry-air component are both a function of the altitude.
See \citet{Schneider:2009hz} for more details on the different
absorption characteristics depending on altitude and wavelength. In any
case, observations at 350 and 450 $\mu$m are possible all year long
for both observatories.

%

\subsection{Sites comparison}

\begin{table*}[!ht]
\caption{\label{transmission_table} First decile and quartiles of the
  350-$\mu$m (top) and 200-$\mu$m (bottom) transmissions for all the
  studied sites.}  \centering
\begin{tabular}{cccccccccc}
\hline Time fraction & Dome C & Dome A & South & Cerro
 & Chaj. & Cerro & Mauna & Summit &
Yangbajing \\ 
2008-2010 & & & Pole & Chaj. & Plat. & Macon & Kea & & \\
\hline
0.10 & 0.62 & 0.72 & 0.61 & 0.65 & 0.56 & 0.49 & 0.40 & 0.42 & 0.19\\ 
0.25 & 0.57 & 0.67 & 0.56 & 0.58 & 0.46 & 0.41 & 0.27 & 0.31 & 0.02\\ 
0.50 & 0.51 & 0.62 & 0.47 & 0.44 & 0.31 & 0.29 & 0.14 & 0.15 & 0.00\\ 
0.75 & 0.41 & 0.57 & 0.34 & 0.24 & 0.12 & 0.15 & 0.03 & 0.02 & 0.00\\ 
\hline 
\hline Time fraction & Dome C & Dome A & South & Cerro
 & Chaj. & Cerro & Mauna & Summit &
Yangbajing \\ 
2008-2010 & & & Pole & Chaj. & Plat. & Macon & Kea & & \\
\hline
0.10 & 0.17 & 0.32 & 0.16 & 0.20 & 0.09 & 0.05 & 0.03 & 0.03 & 0.00\\ 
0.25 & 0.11 & 0.22 & 0.11 & 0.12 & 0.04 & 0.02 & 0.01 & 0.01 & 0.00\\ 
0.50 & 0.07 & 0.16 & 0.05 & 0.04 & 0.01 & 0.00 & 0.00 & 0.00 & 0.00\\ 
0.75 & 0.03 & 0.11 & 0.01 & 0.00 & 0.00 & 0.00 & 0.00 & 0.00 & 0.00\\
\end{tabular}
\end{table*}

Based upon the method described in \citet{Tremblin:2011fg}, we use MOLIERE
to calculate the transmissions at 200 $\mu$m and 350 $\mu$m. We do not
compute here the transmissions for SOFIA because it is by far superior
compared to all ground based sites. Its transmission curves will be discussed in a dedicated
section (Sect.~5).

\begin{figure}[!ht]
\centering \includegraphics[width=\linewidth]{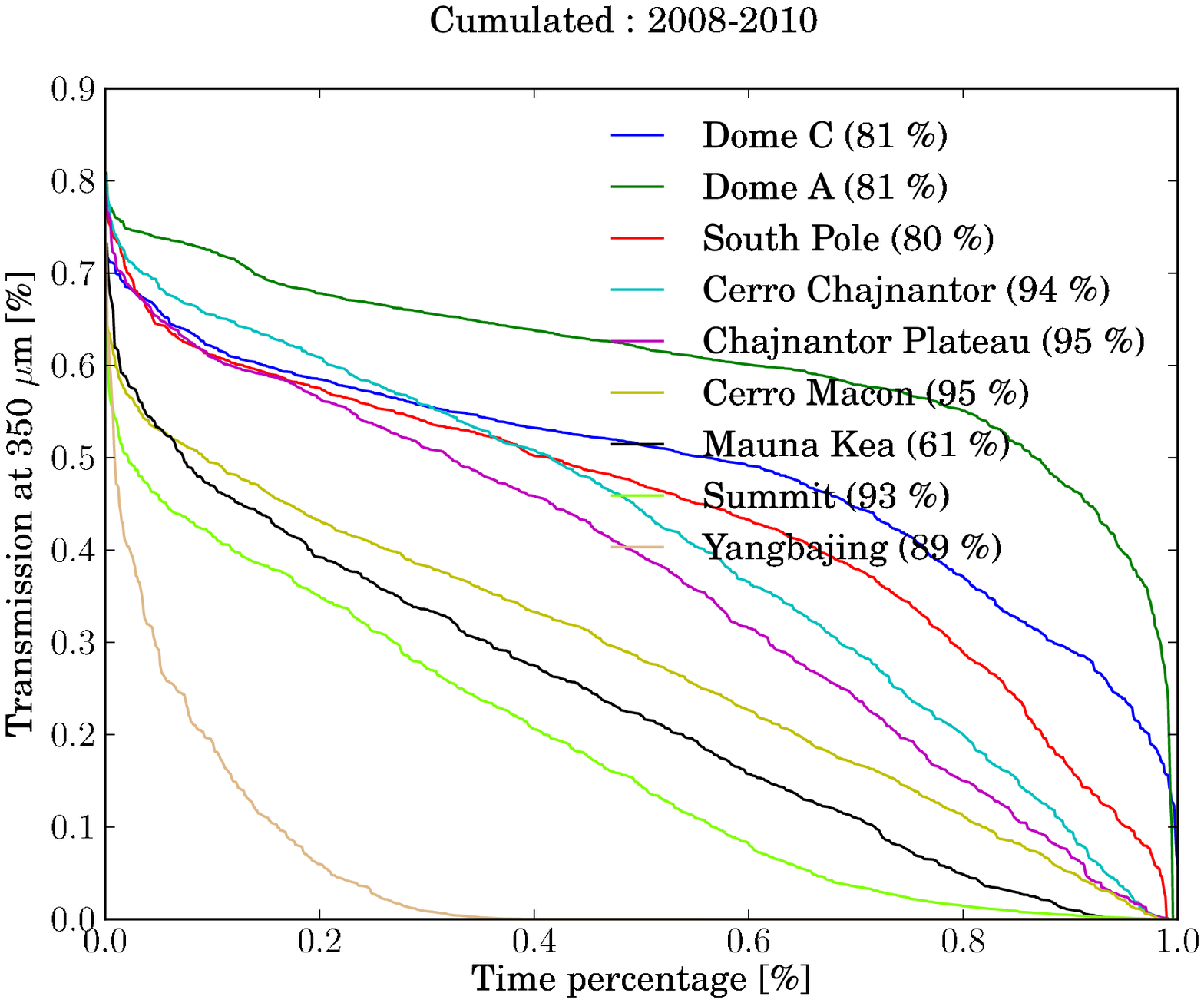}
\includegraphics[width=\linewidth]{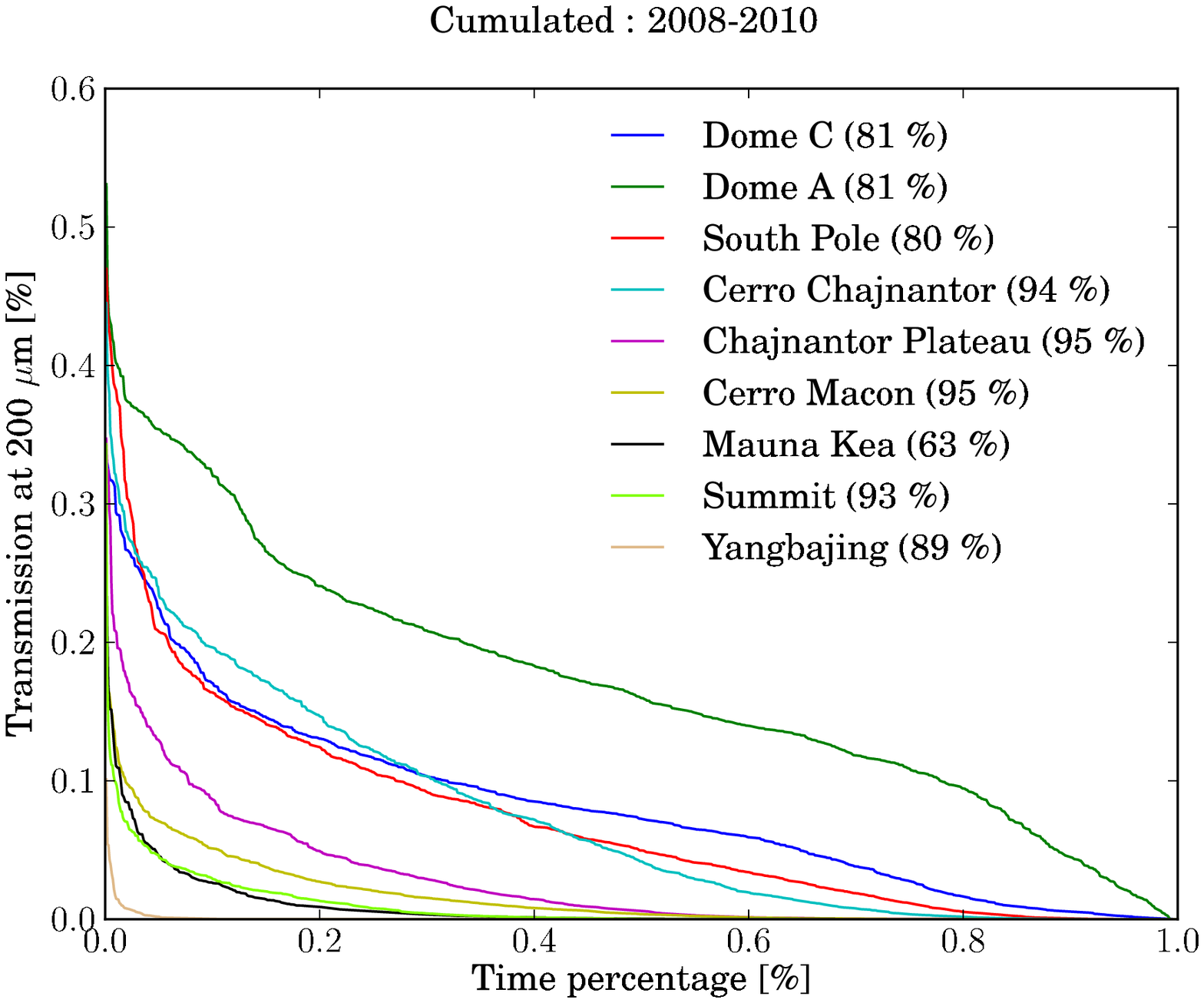}
\caption{Comparison of the functions of repartition of the 350-$\mu$m
  (top) and 200-$\mu$m (bottom) transmissions for all the sites
  between 2008 and 2010. The x-axis time percentage is done over the
  period on which the data extraction for the sites was possible, the
  actual percentage of data we got is indicated for each site in
  parenthesis. }
\label{transmission_comp}
\end  {figure}

Figure \ref{transmission_comp} shows the functions of repartition of
the 350- and 200- $\mu$m transmissions for all the other sites. In
Table \ref{transmission_table}, we give the first decile and the
quartiles of 200- and 350- $\mu$m transmissions. 25 \% of the time, a
transmission better than 55-65 \% (10-20 \%) at 350 $\mu$m (200
$\mu$m) can be found in Antarctica. For the first quartile, Cerro
Chajnantor catches up the Antarctic sites thanks to its high
altitude. However for the second and third quartiles Antarctic sites
have a better transmissions based on the long term statistics, especially at
350 $\mu$m. For the three sites, Chajnantor Plateau, Cerro Macon and
Mauna Kea, the transmission window at 350 $\mu$m opens significantly
while it is only rarely possible to observe at 200 $\mu$m.


The extreme stability of the Antarctic sites is clearly visible in
Fig. \ref{transmission_comp}. There is a stable plateau for the
transmission between the first and third quartiles that allows long
time series and surveys to be conducted on these sites. This is the
direct consequence of the stability of the PWV that was previously
identified at Dome C in Fig. \ref{domec_pwv}. With its higher
altitude, Dome A presents the best conditions on Earth for
submillimetre astronomy in terms of transmission and day-to-day
stability.

\subsection{Transmission stability at 200 $\mu$m}

\begin{figure}[!ht]
\centering \includegraphics[width=\linewidth]{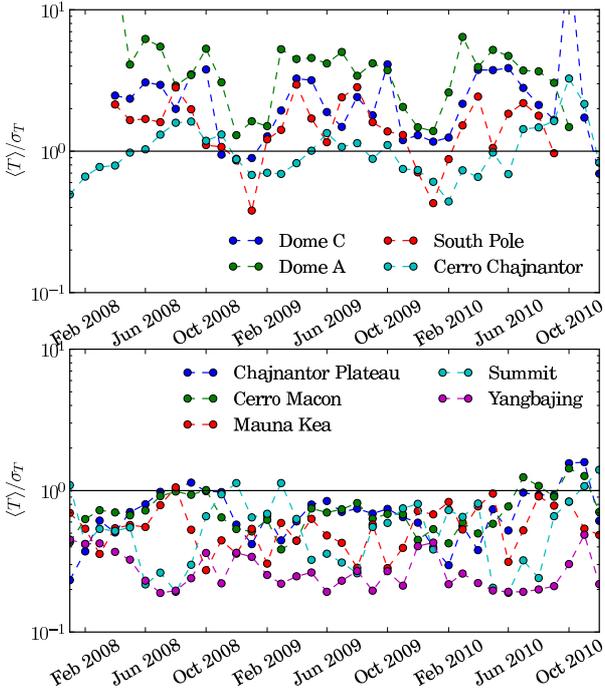}
\caption{SPQR ratio for the 200-$\mu$m transmission for all the sites between 2008 and 2010. The mean
  values between 2008 and 2010 are Dome A 3.6, Dome C 2.7, South Pole
  1.3, Cerro Chajnantor 1.1, Chajnantor Plateau 0.7, Cerro Macon 0.7,
  Mauna Kea 0.6, Summit (Greenland) 0.6, Yangbajing (Tibet) 0.3.}
\label{spqr_ratio}
\end  {figure}

The stability of the transmission is an important parameter for large
surveys and time-series studies. The PWV of Chilean Sites is highly
variable from one day to another whereas Antarctic sites are much more
stable (see Figs. \ref{domec_pwv} and \ref{ccat_pwv}). The stability
of the transmission leads to a high transmission even at long time
fraction in the repartition functions of
Fig. \ref{transmission_comp}. In order to quantify and compare the
stability, we used the indicator introduced by
\citet{DeGregori:2012hx}, the site photometric quality ratio
(SPQR). It consists in the ratio of the monthly averaged
transmission to its monthly standard deviation, on a daily
time-scale
\begin{equation}
\mathrm{SPQR} = \langle T \rangle / \sigma_T
\end{equation}
The variations of this ratio for the transmission at
200 $\mu$m between 2008 and 2010 are
plotted in Fig. \ref{spqr_ratio} and indicate 
that all temperate sites have a SPQR ratio lower than 1 while
Antarctic sites have a ratio greater than 1. Therefore, the
fluctuations of the transmission are greater than the averaged
transmission, on temperate sites. This quantifies the high variability
of the transmission at these sites. Note that the Arctic site on the
Summit mountain is also highly variable, hence Antarctica is really
unique even among polar environments. On the Antarctic
plateau, Dome A has the best SPQR ratio with a monthly-averaged
transmission that is typically 3-4 times higher than the
fluctuations. Dome C achieves also very good conditions with a ratio
of the order of 2-3 while the South Pole has a monthly-averaged
transmission of the same order of the fluctuations comparable to the
conditions reached at Cerro Chajnantor. 

%
%

\section{Atmospheric transmission for SOFIA}

The first successfull science flights with SOFIA were carried out in
2011 and summarized in special issues of A\&A (2012, Vol. 542,
spectroscopic observations with GREAT) and ApJ (2012, Vol. 749,
continuum imaging). In particular the spectroscopic observations at
high frequencies (e.g. the [CII] line at 1.9 THz) showed that although
the transmission is very high at these frequencies (typical PWVs are
in the range of a few $\mu$m up to 50 $\mu$m for altitudes between 11
and 14 km), the calibration of the data using atmospheric models is
not straightforward and requires a more detailed understanding of the
atmospheric processes at the troposphere/stratosphere border. As
pointed out in \citet{Guan:2012ce}, atmospheric models are particularly
important because atmospheric absorption lines are more narrow at
these altitudes and their small-scale variation is very difficult to
measure. In \citet{Guan:2012ce}, three atmospheric models are compared: AM
\citep{Paine:2011}, ATRAN (Lord 1992, priv. comm.) and MOLIERE
\citep{Urban:2004cr}. Special emphasize is given separately to the
wet- and 
dry-component. While all models agree well for the wet component,
ATRAN does not include the collision-induced absorption by N$_2$ and
O$_2$, and AM calculates a two times lower opacity than
MOLIERE. Results of versatile models like MOLIERE depend much on
choices made by the user. At
this stage, it is not clear which models fits best for the data
calibration and studies to compare the AM and MOLIERE codes are under
way.

\begin{figure}[!ht] 
\centering \includegraphics[width=0.6\linewidth,angle=-90]{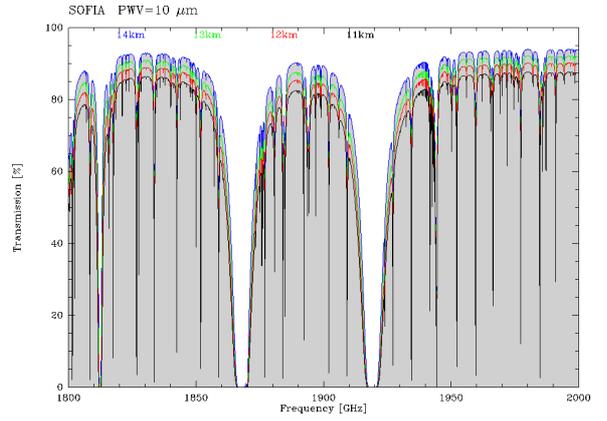}
\caption{ Atmospheric transmission curve for 4 flight altitudes of
  SOFIA calculated with MOLIERE at a PWV of 10 $\mu$m. }
\label{sofia_11-14}
\end  {figure}

\begin{figure}[!ht] 
\centering \includegraphics[width=0.6\linewidth,angle=-90]{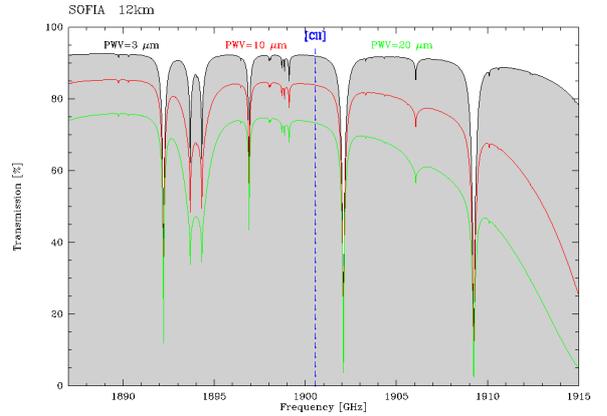}
\caption{Zoom into the transmission curve for the frequency range
  containing the [CII] 1.9 THz for 3 different PWV at an altitude of
  12 km calculated with MOLIERE. The [CII] line is indicated by a
  dashed blue line.}
\label{sofia_cii}
\end  {figure}

In this paper, we present selected model results of MOLIERE calculated
for 4 different altitudes (11 to 14 km in steps of 1 km), three PWVs
(3, 10 and 20 $\mu$m), and a frequency range of 1800 to 2000 GHz. We
chose this frequency range because it includes the astronomically
very important atomic finestructure emission line of ionized carbon at
1900 GHz. Figure~\ref{sofia_11-14} shows the transmission curves for
the 4 altitudes at a PWV of 10 $\mu$m. Generally, the transmission is
high ($>$80 \%) except of the deep and broad water absorption features
and line absorption of minor species. Only in these ranges, the
atmosphere is opaque even at those high altitudes.
Figure~\ref{sofia_cii} shows a zoom into the frequency range around
the [CII] 1.9 THz line for different PWVs. The transmission remains
high ($\sim$75 \%) even for a PWV of 20 $\mu$m.

%
%

\section{Summary and conclusions}

We presented PWV statistics and the corresponding transmission curves
for potential and existing submillimetre observational sites all around the
world. Thanks to satellite measurements (IASI) it is possible for the
first time to conduct a comparison which is not biased by the
different calibration techniques of ground-based instruments that have
very different behaviors caused by their working conditions. This is
particularly true for comparison between polar and non-polar sites
because of the harsh environmental conditions at the poles.

Comparisons were done for the PWV statistics and for the transmissions
using the atmospheric model MOLIERE. For astronomical purposes, a
comparison using the transmissions is mandatory since the altitude of
the site has also a strong influence on the resulting
transmission. For a percentage of 25 \% of the year, Cerro Chajnantor
presents very good transmissions that are comparable to Antarctic
sites. However on the long term, only Antarctic sites provide a stable
transmission that will allow unique science cases to be studied
there. Observations at 350 $\mu$m and 450 $\mu$m are possible all year
long while the 200 $\mu$m window opens significantly for at least 25
\% of the year. Furthermore, Antarctic sites are free of day-night
variations thanks to the long polar night. All the other sites studied
here have a highly-variable transmission. The
typical monthly-averaged transmission at 200 $\mu$m is lower than its
fluctuations on all the sites except for Dome A and Dome C. Therefore, 
these two sites will be unique for surveys and time-series studies in the
submm range. 

The method used to compare the different sites is robust and based on
only one instrument, IASI, and the atmospheric model
MOLIERE. We derived in this paper the PWV and atmospheric transmission on
  well-known sites for submillimetre astronomy
  and showed that it is possible to retrieve statistics that are in a
  good agreement with in-situ measurements. A
calculator to show these PWV statistics and to compute the corresponding
transmission at any given wavelength is available to the community\footnote{\url{http://irfu.cea.fr/submm} and
\url{http://submm.eu}} for all the sites
presented here and for the three year 2008, 2009 and 
2010. The year 2011 will be added to the calculator. Other
  potential sites could be investigated upon request\footnote{Contact:
    pascal.tremblin@cea.fr}.\\

In conclusion, this method could identify sites on Earth with a great
potential for submillimetre astronomy, and guide future site testing
campaigns \it{in situ}.

%
%

\begin{acknowledgements}
We thank the anonymous referee for his/her comments that helped to improve 
this paper, in particular pointing out the importance of the pressure level to retrieve
correct PWVs. 
\end{acknowledgements}

%
%

\bibliographystyle{aa} \bibliography{main.bib}

%
%

\appendix \label{appendix}
\section{PWV statistics and transmission curves}

\begin{figure}[!ht]
\centering \includegraphics[width=0.8\linewidth]{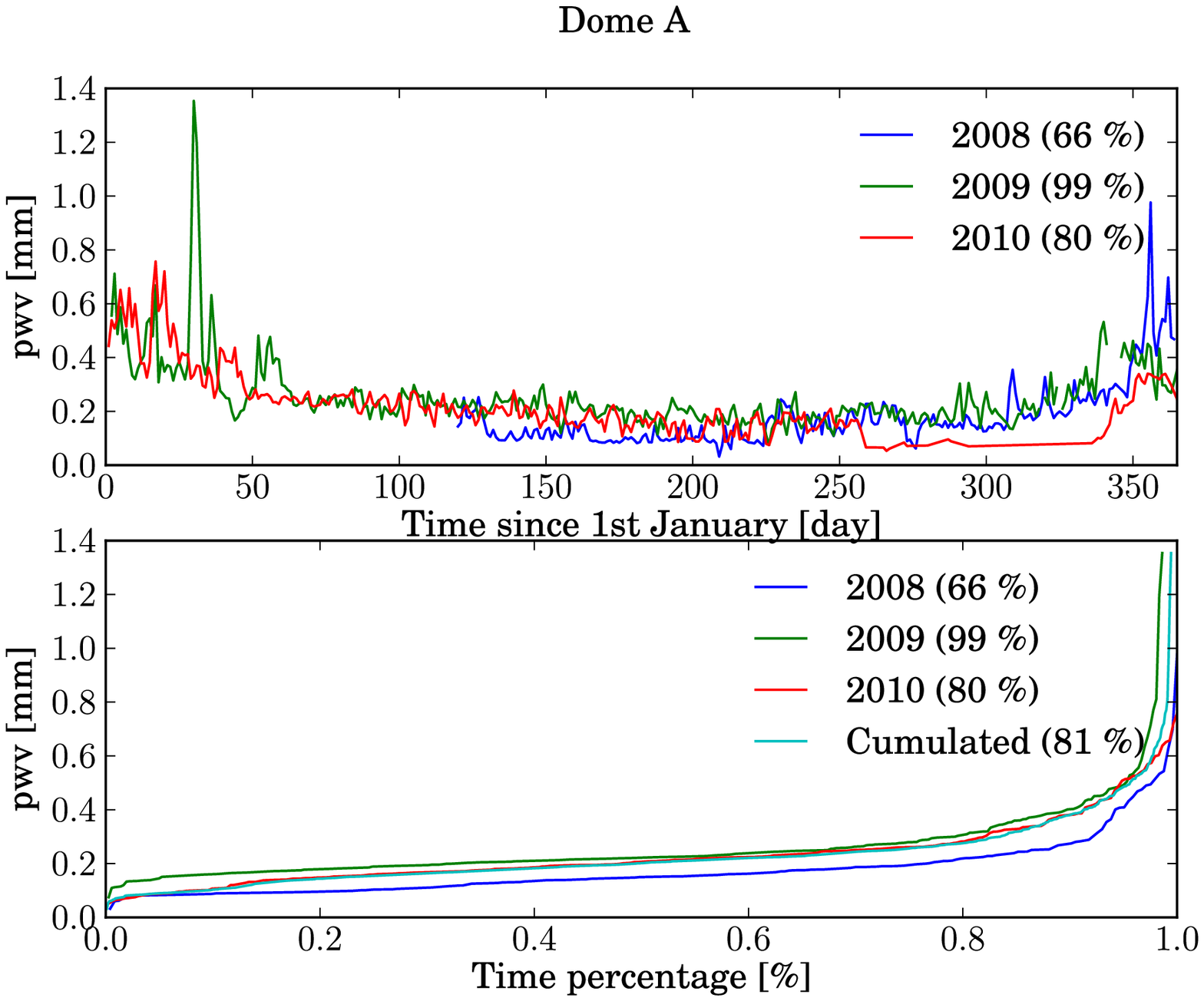}
\includegraphics[width=0.48\linewidth,angle=-90]{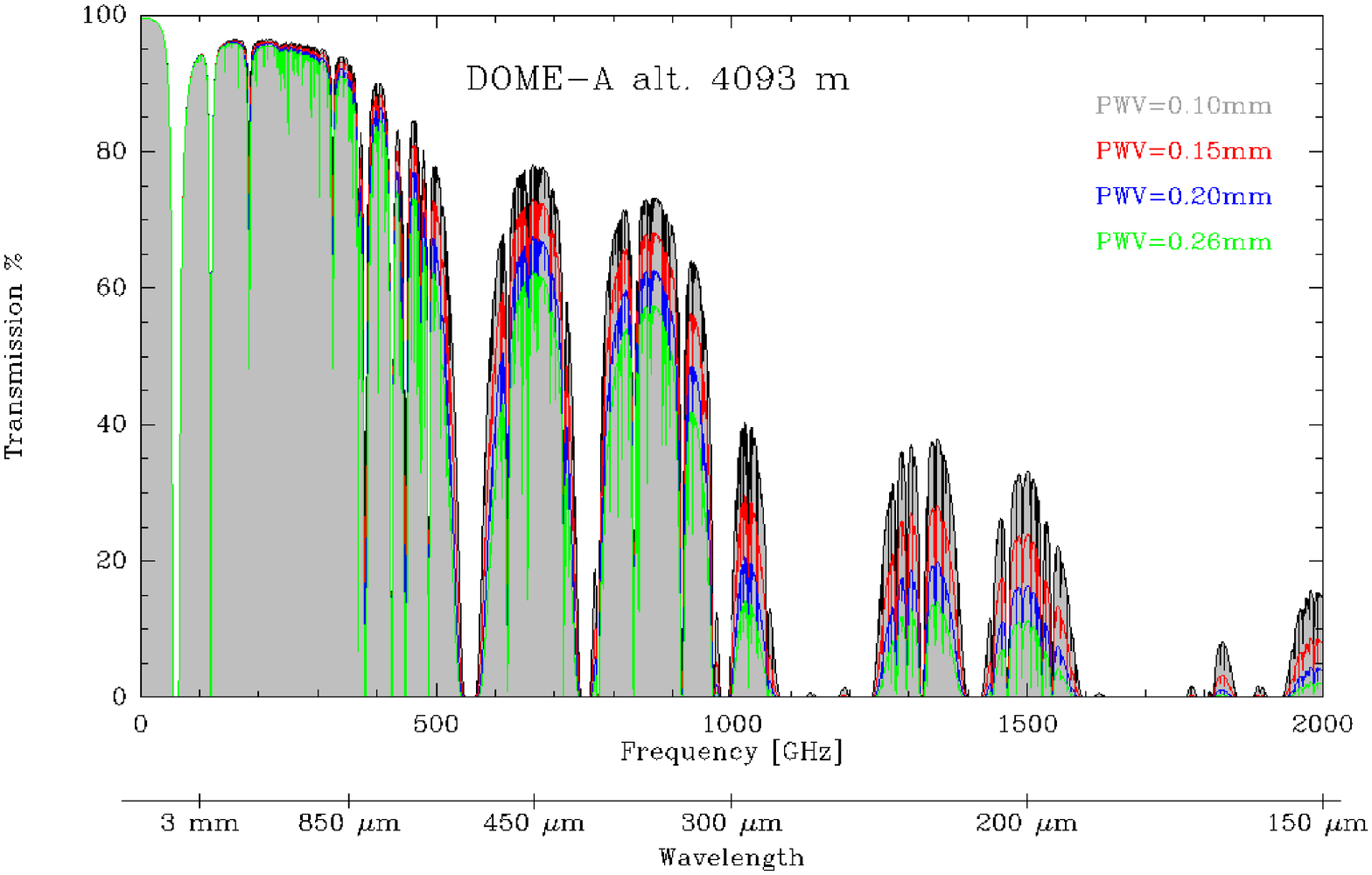}
\caption{PWV statistics (top) and transmission curves (bottom) for
  Dome A. The transmission curve for the first decile of PWV is in
  grey and the quartiles of PWV are given by red: 25 \%, blue: 50 \%,
  green: 75 \%.}
\end  {figure}
\begin{figure}[!ht]
\centering \includegraphics[width=0.8\linewidth]{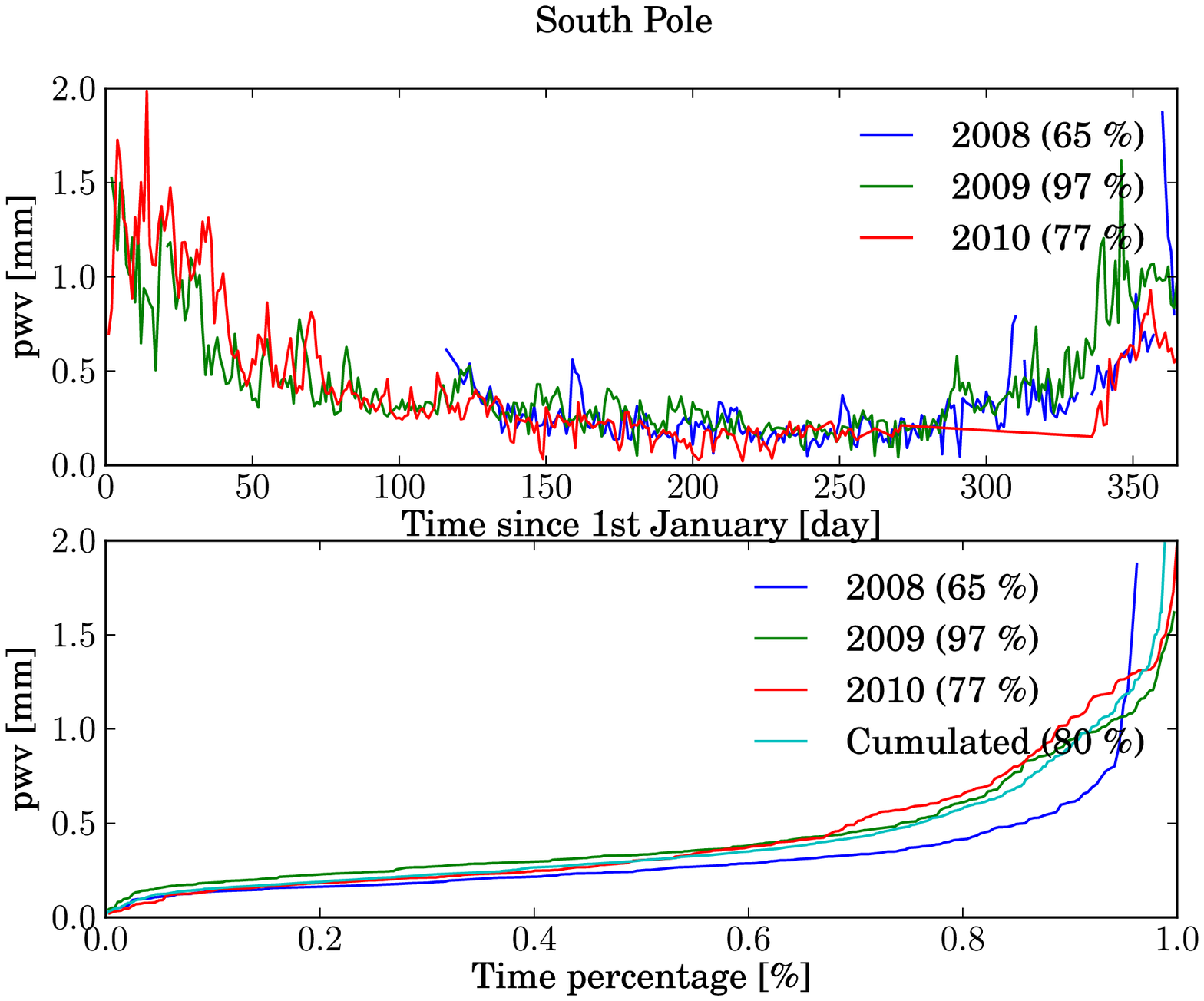}
\includegraphics[width=0.48\linewidth,angle=-90]{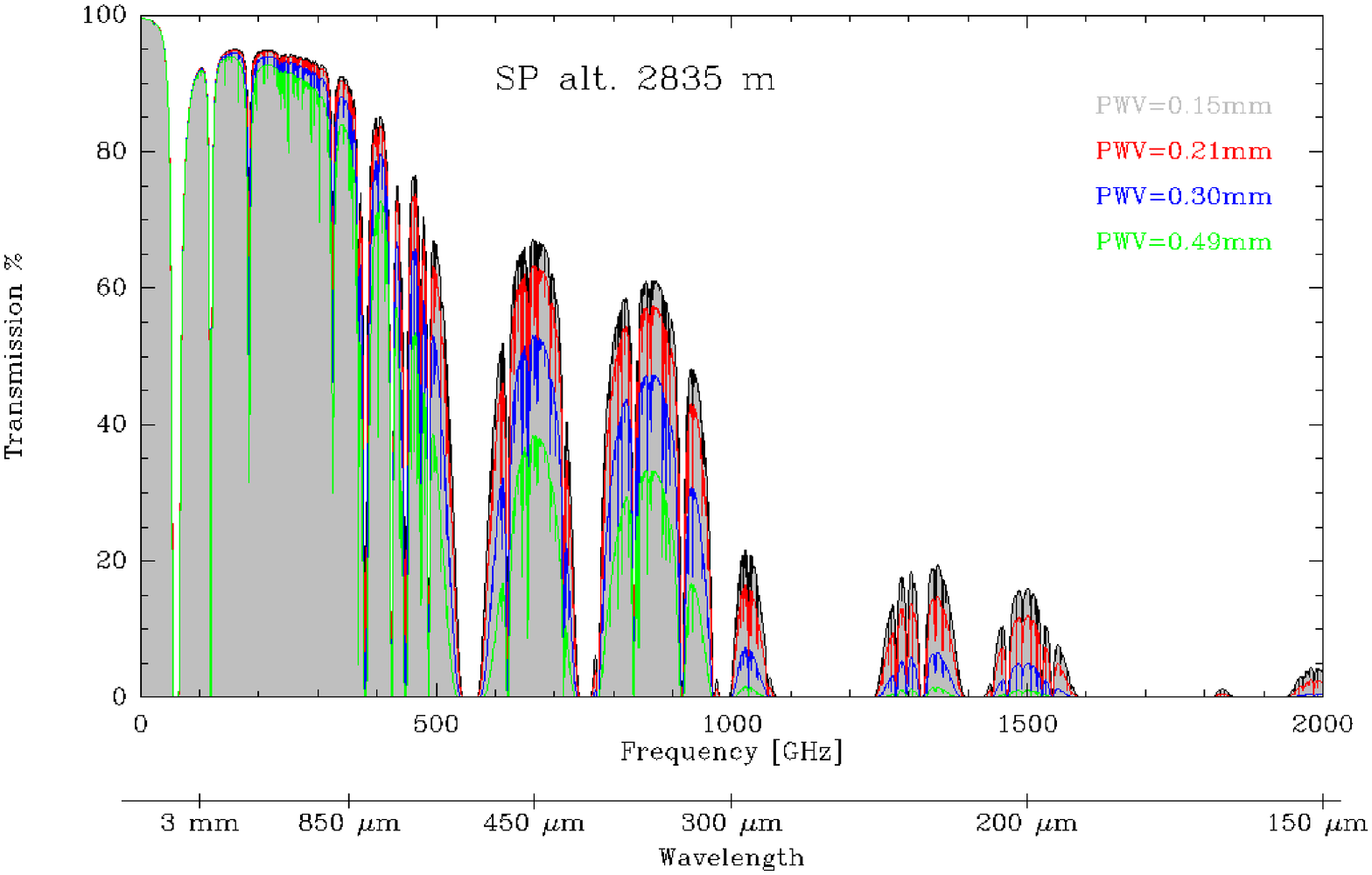}
\caption{PWV statistics (top) and transmission curves (bottom) for the
  South Pole. The transmission curve for the first decile of PWV is in
  grey and the quartiles of PWV are given by red: 25 \%, blue: 50 \%,
  green: 75 \%. }
\end  {figure}
\begin{figure}[!ht]
\centering \includegraphics[width=0.8\linewidth]{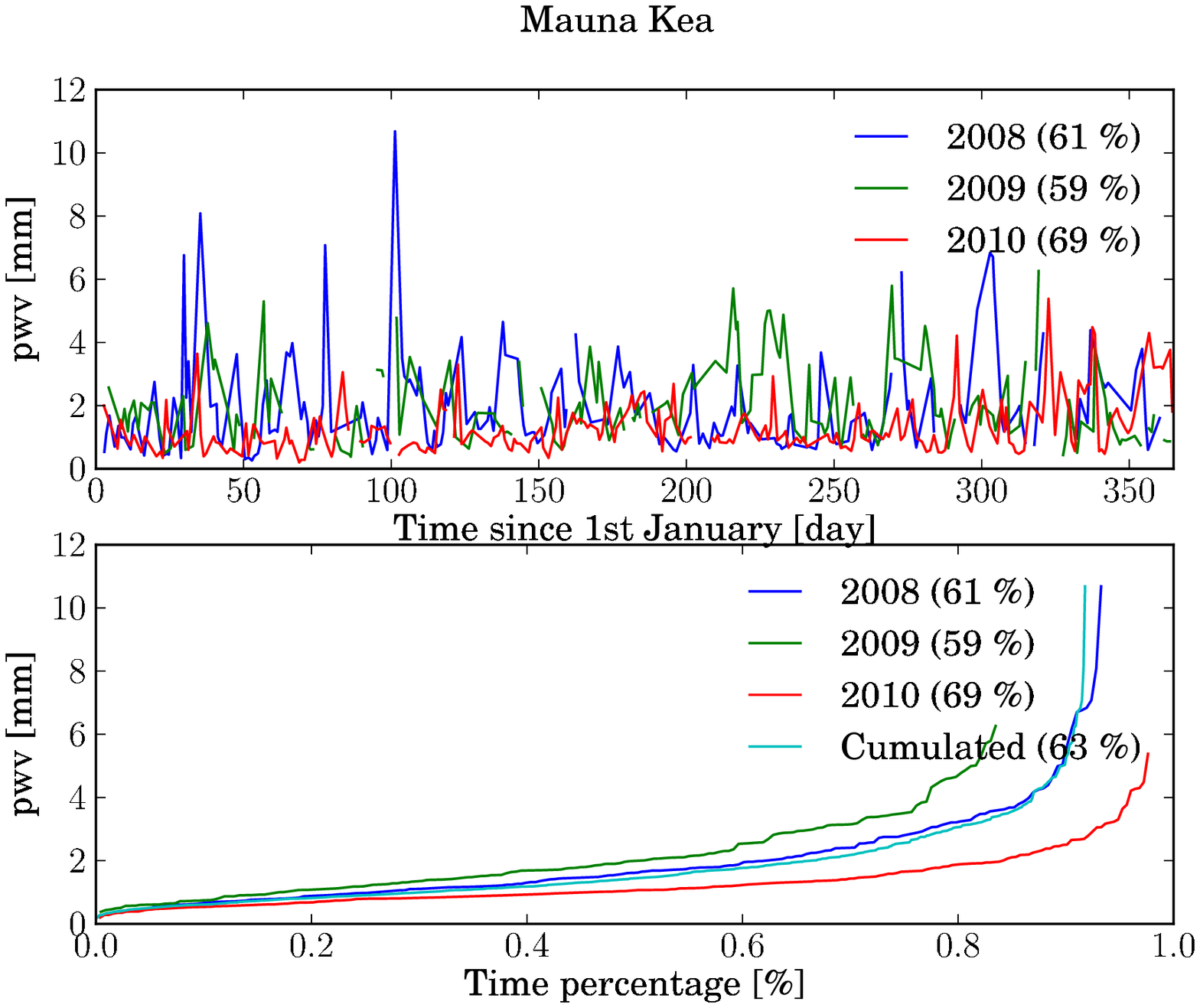}
\includegraphics[width=0.48\linewidth,angle=-90]{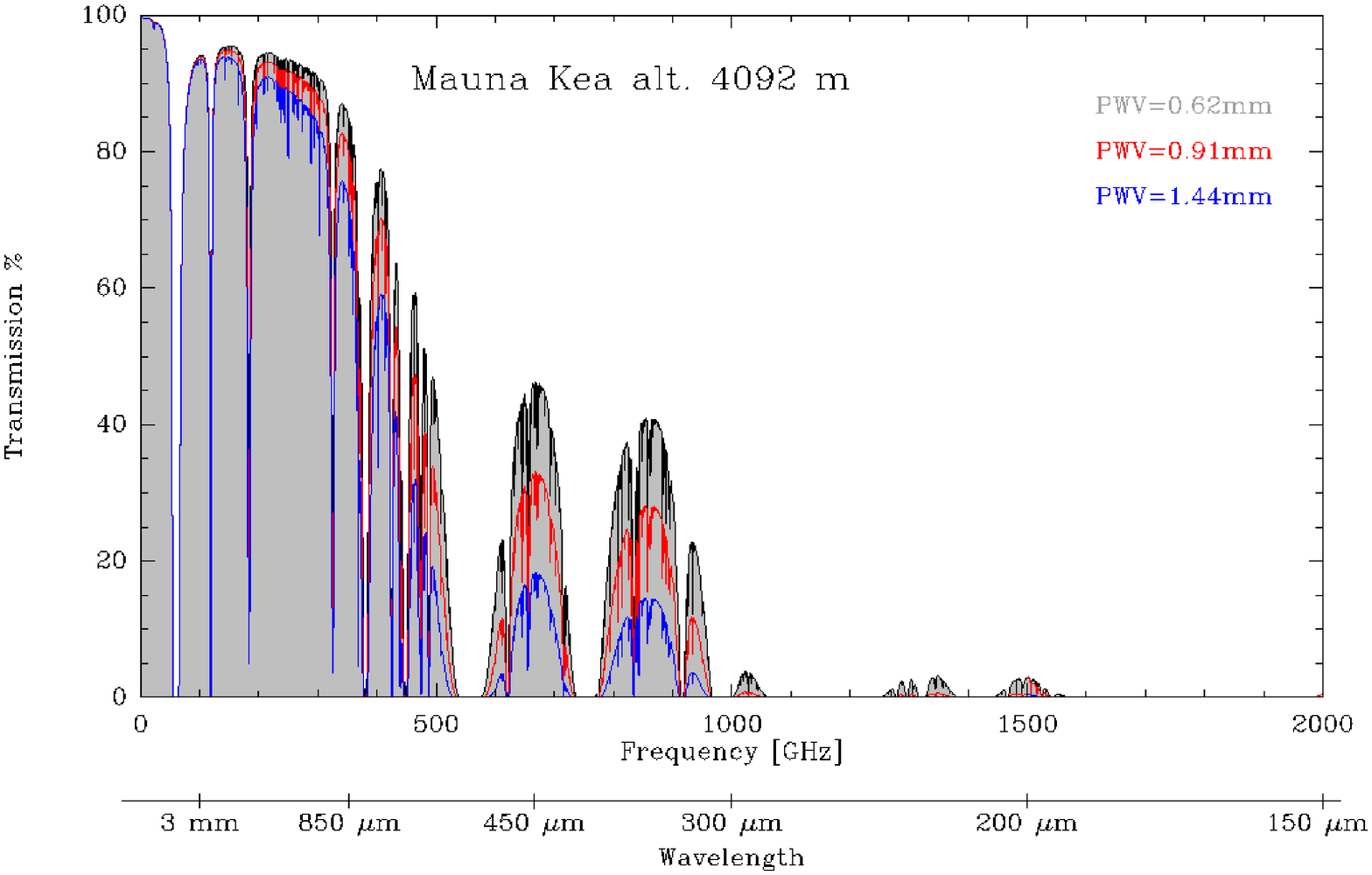}
\caption{PWV statistics (top) and transmission curves (bottom) for
  Mauna Kea. The transmission curve for the first decile of PWV is in
  grey and the quartiles of PWV are given by red: 25 \%, blue: 50 \%.}
\end  {figure}
\begin{figure}[!ht]
\centering \includegraphics[width=0.8\linewidth]{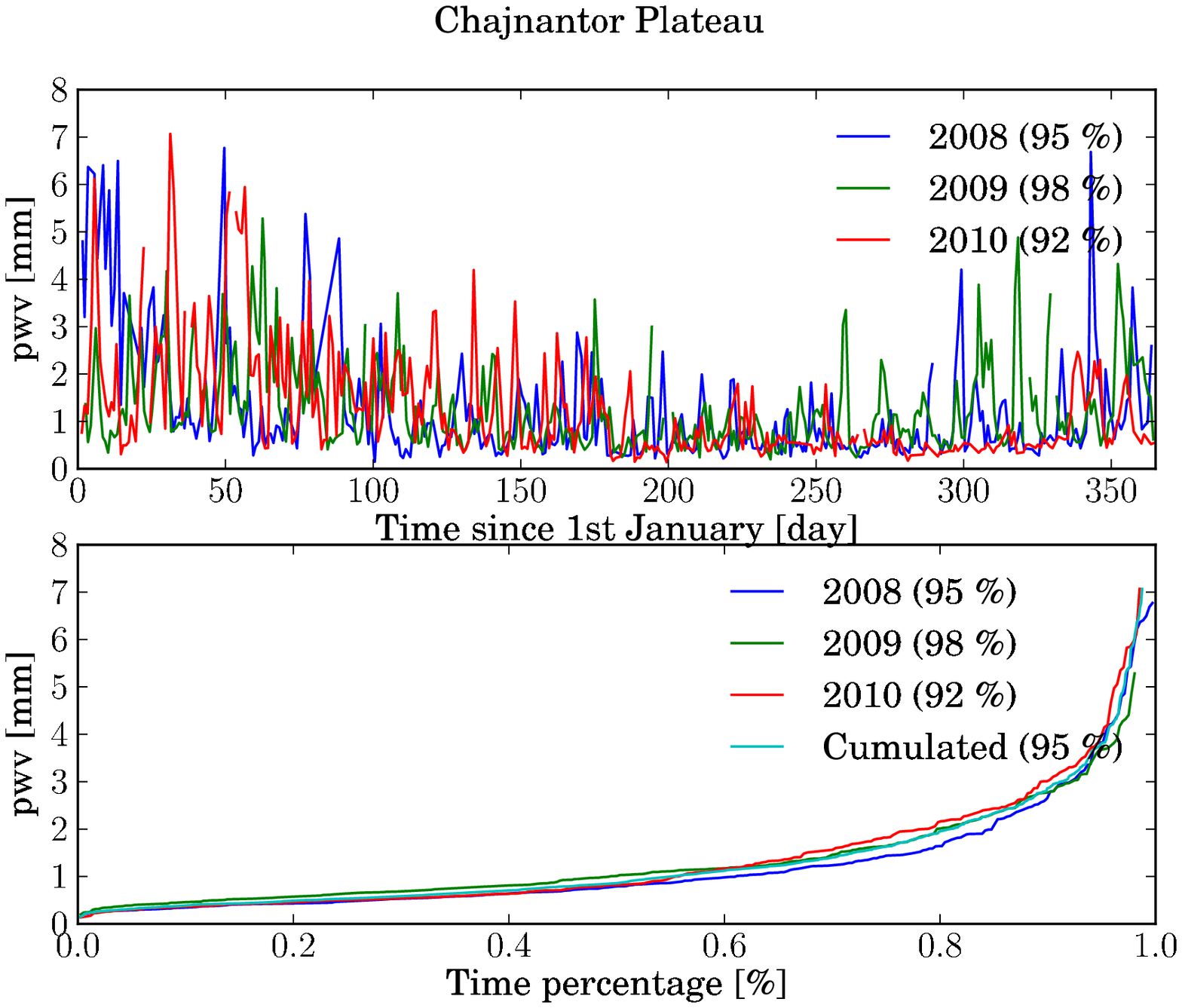}
\includegraphics[width=0.48\linewidth,angle=-90]{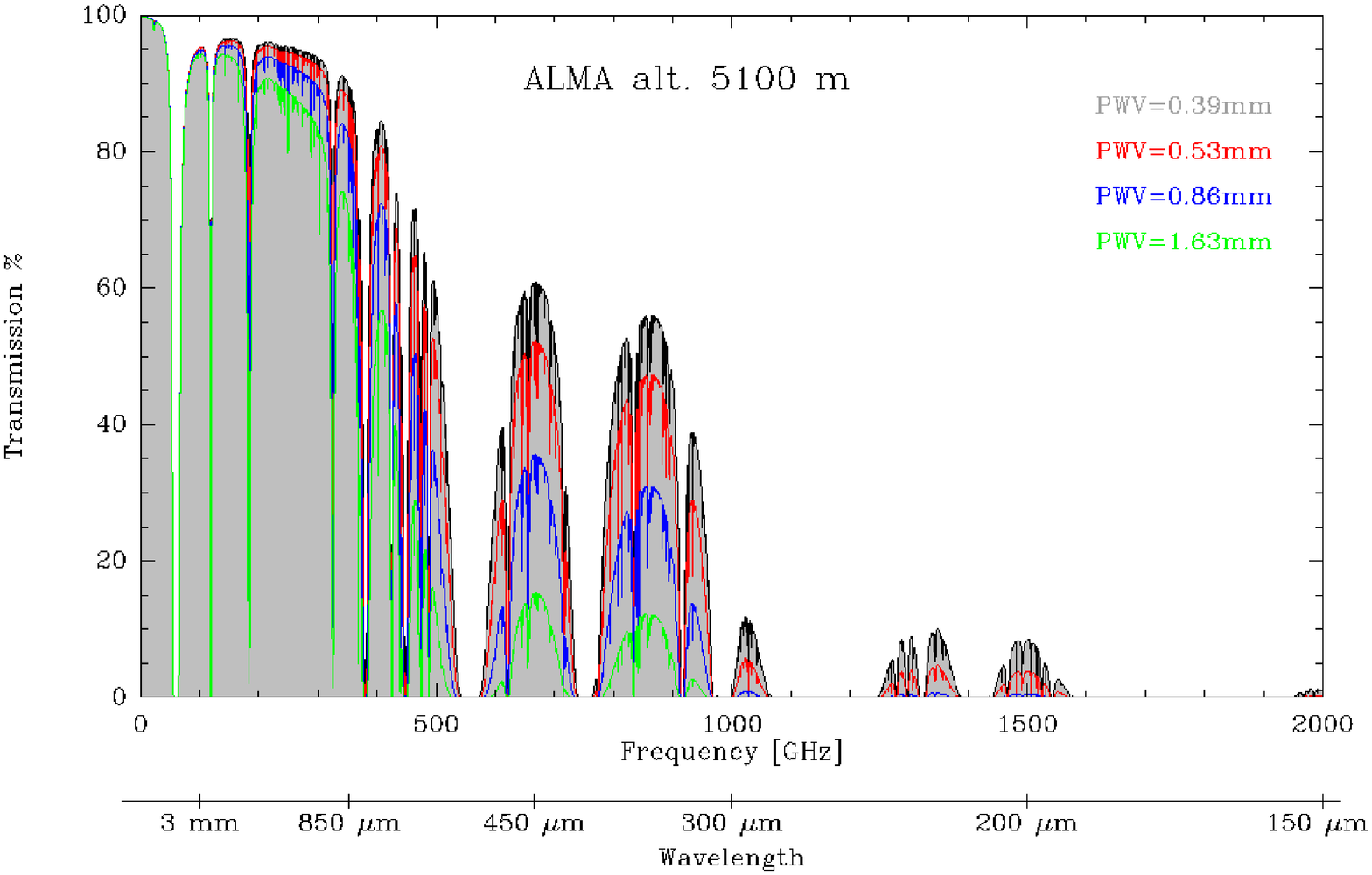}
\caption{PWV statistics (top) and transmission curves (bottom) for
  Chajnantor Plateau. The transmission curve for the first decile of
  PWV is in grey and the quartiles of PWV are given by red: 25 \%,
  blue: 50 \%.}
\end  {figure}
\begin{figure}[!ht]
\centering \includegraphics[width=0.8\linewidth]{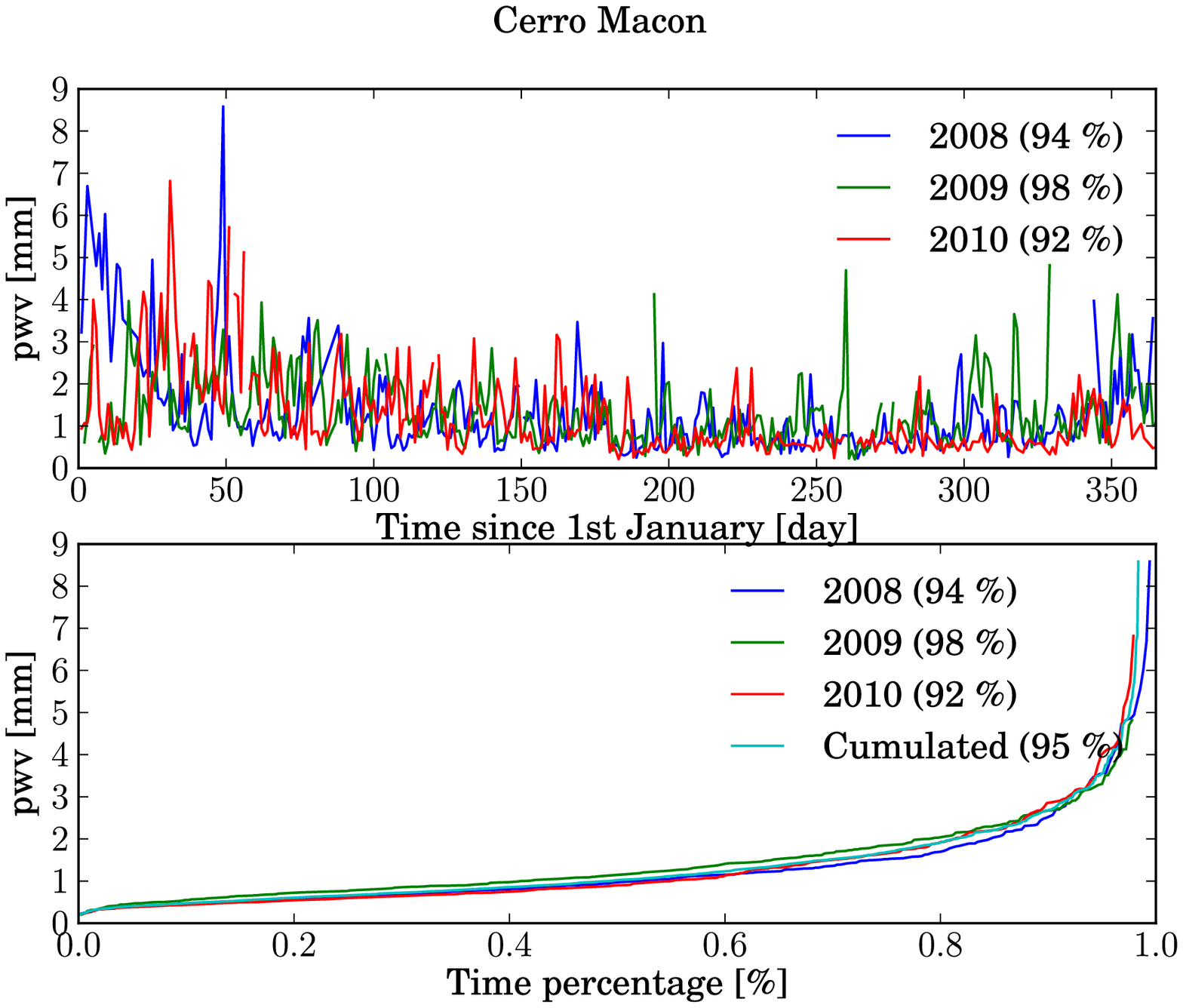}
\includegraphics[width=0.48\linewidth,angle=-90]{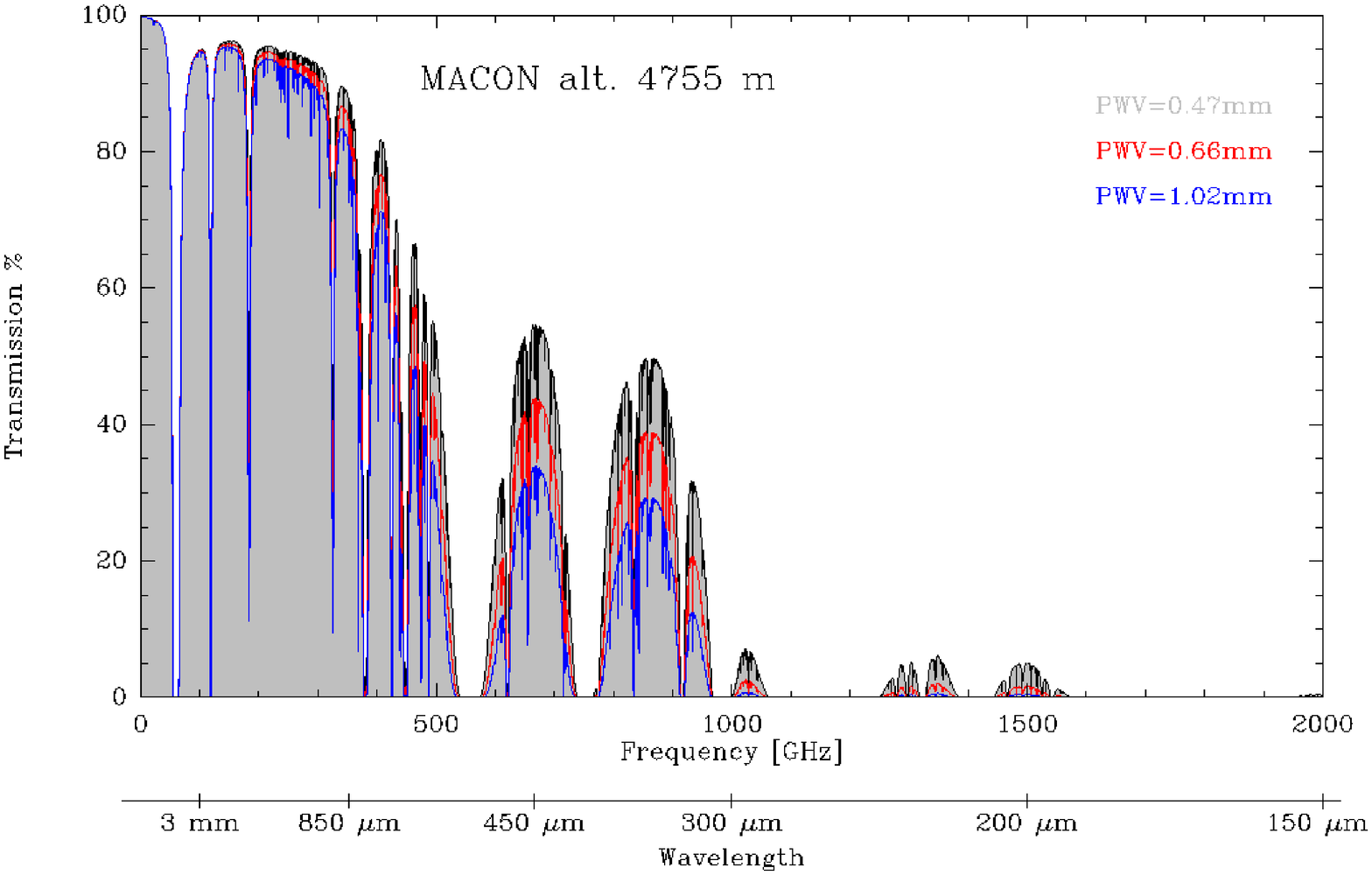}
\caption{PWV statistics (top) and transmission curves (bottom) for
  Cerro Macon. The transmission curve for the first decile of PWV is
  in grey and the quartiles of PWV are given by red: 25 \%, blue: 50
  \%.}
\end  {figure}
\begin{figure}[!ht]
\centering \includegraphics[width=0.8\linewidth]{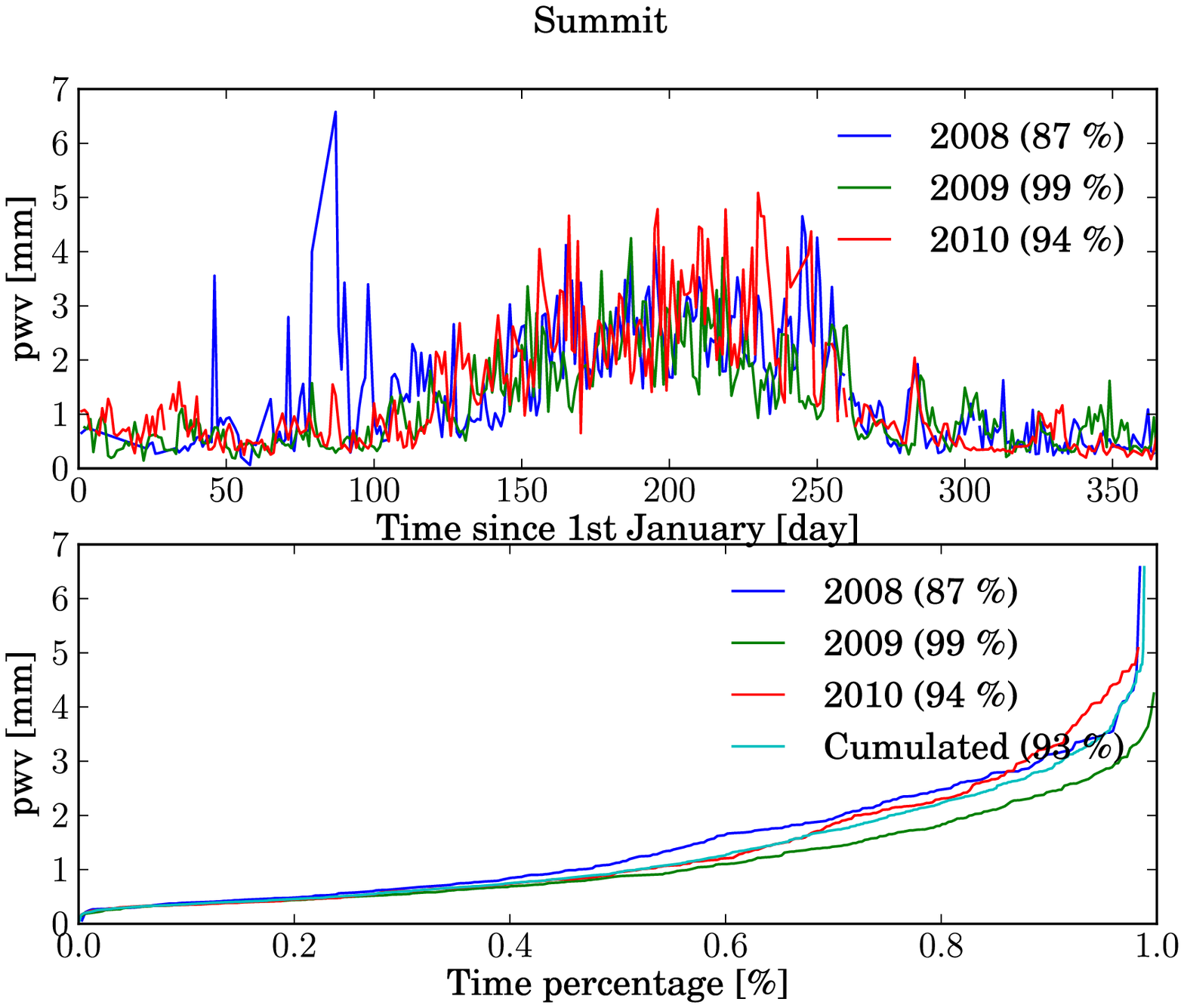}
\includegraphics[width=0.48\linewidth,angle=-90]{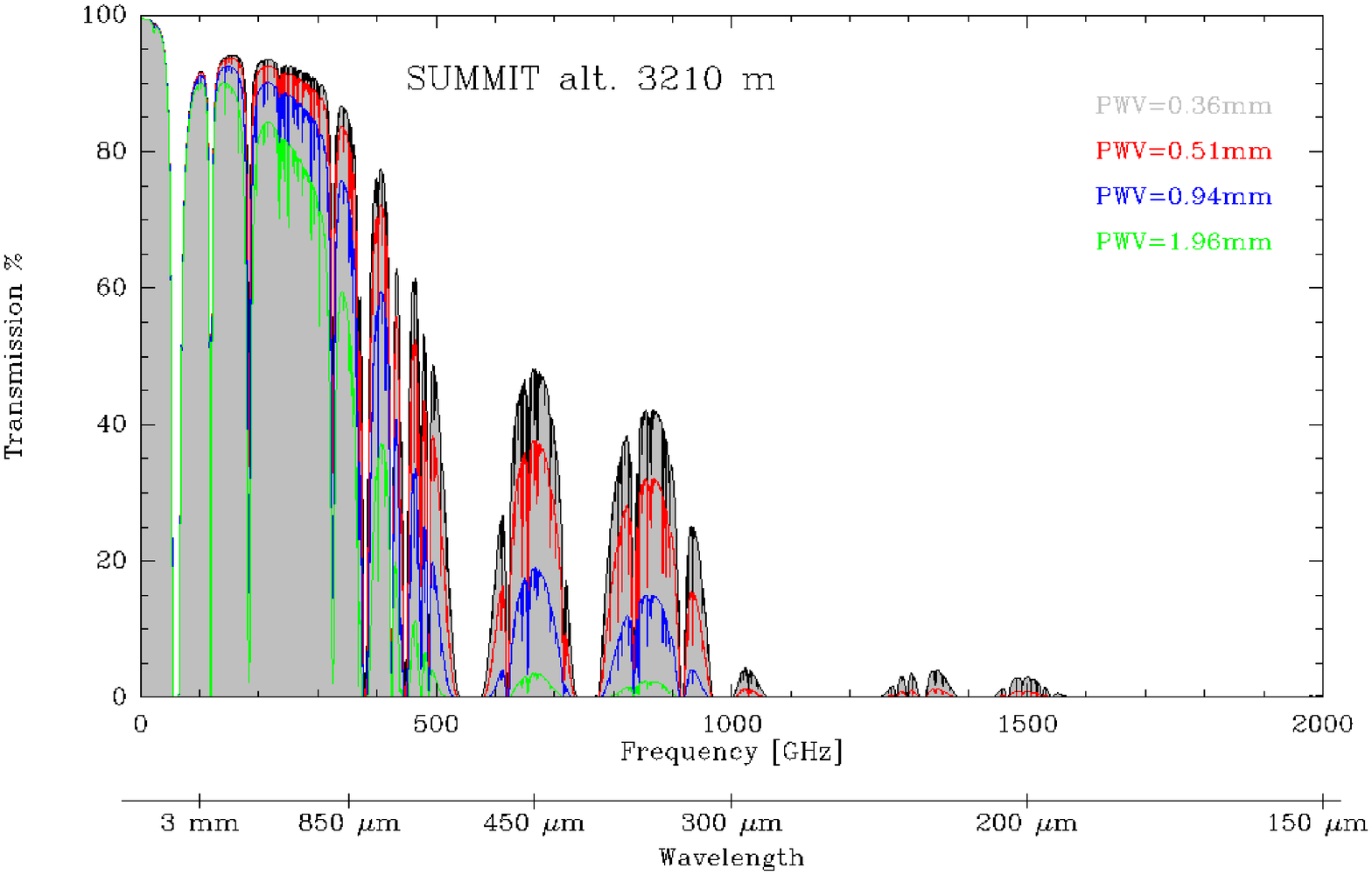}
\caption{PWV statistics (top) and transmission curves (bottom) for
  Summit (Greenland). The transmission curve for the first decile of PWV is in
  grey and the quartiles of PWV are given by red: 25 \%, blue: 50 \%,
  green: 75 \%.}
\end  {figure}
\begin{figure}[!ht]
\centering \includegraphics[width=0.8\linewidth]{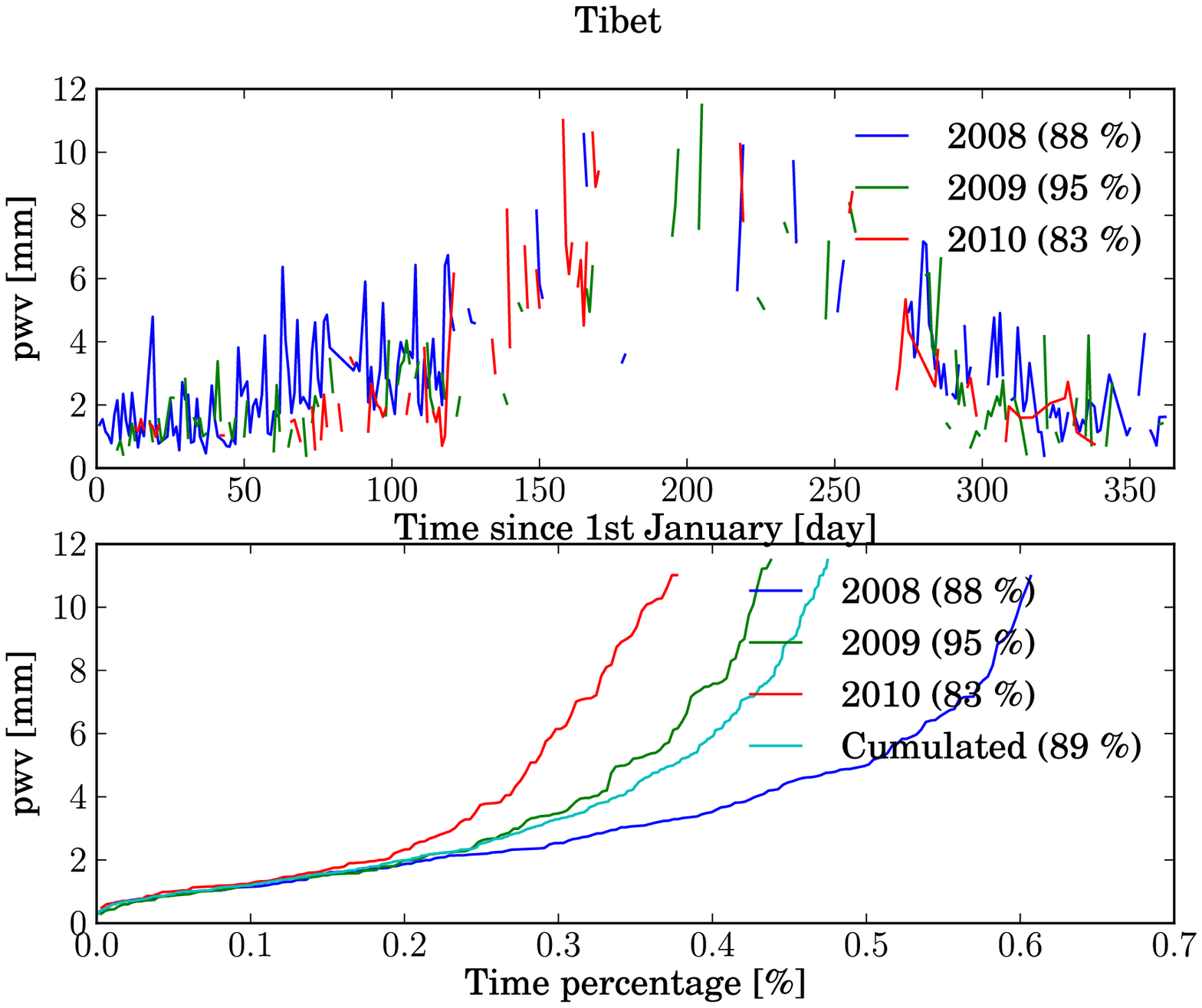}
\includegraphics[width=0.48\linewidth,angle=-90]{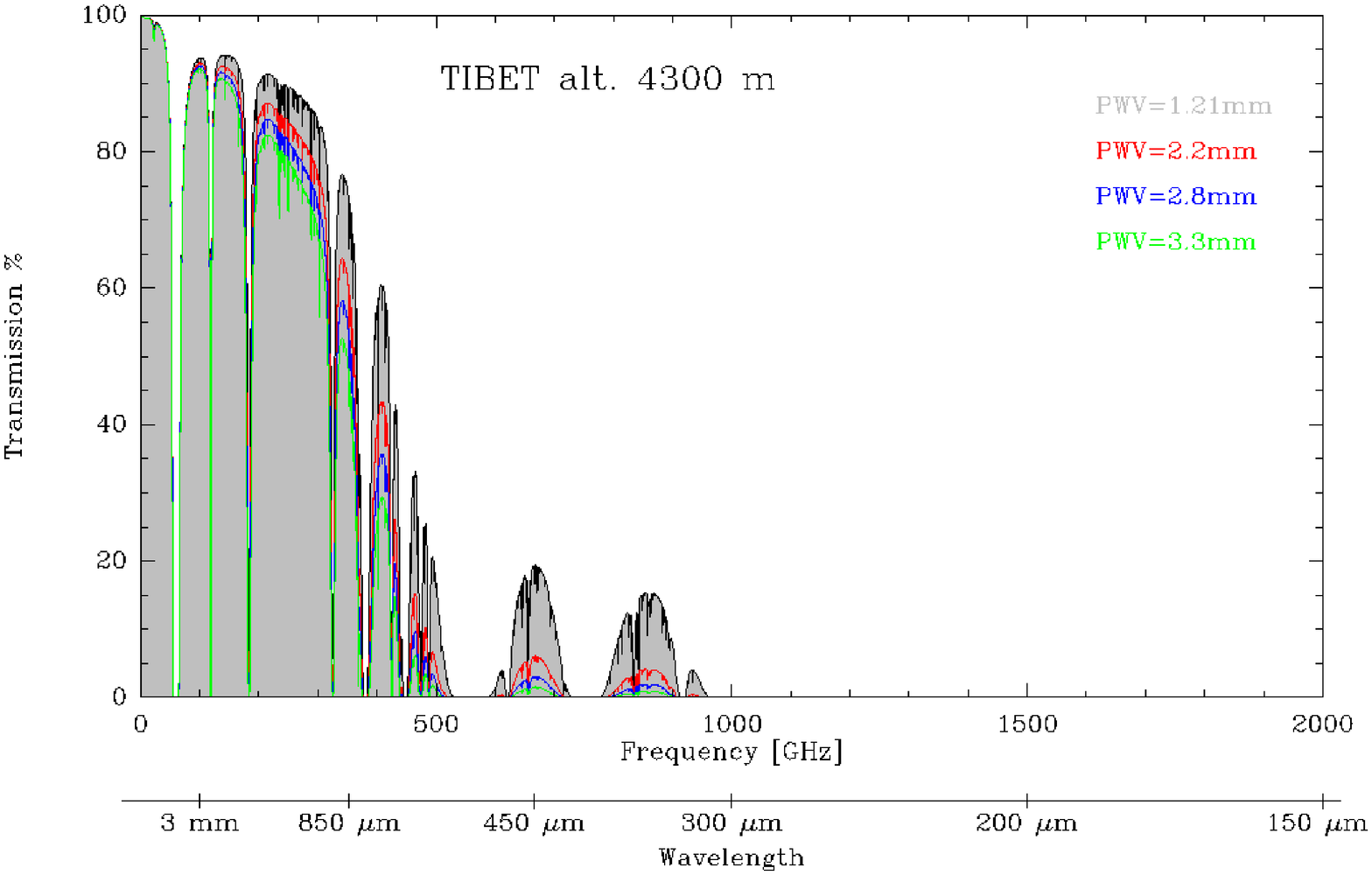}
\caption{PWV statistics (top) and transmission curves (bottom) for
  Yangbajing (Tibet). The transmission curve for the first decile of PWV is in
  grey and the quartiles of PWV are given by red: 25 \%, blue: 50 \%,
  green: 75 \%.}
\end  {figure}
\begin{figure}[!ht]
\centering \includegraphics[width=0.8\linewidth]{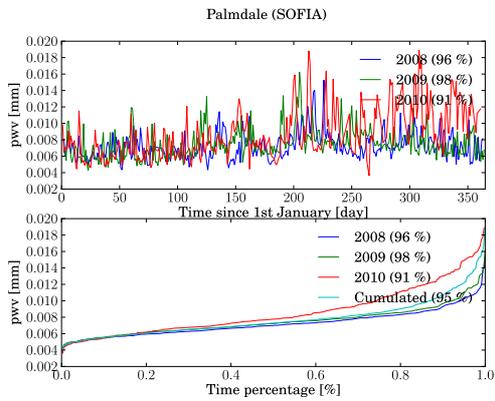}
\caption{PWV statistics for Palmdale USA, at a stratospheric
  altitude of 12km.}
\end  {figure}
\begin{figure}[!ht]
\centering \includegraphics[width=0.8\linewidth]{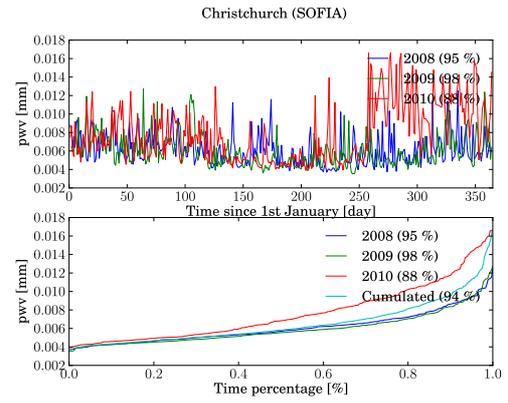}
\caption{PWV statistics for Christchurch New Zealand, at a
  stratospheric altitude of 12 km.}
\end  {figure}

\end {document}